\documentclass{aa}  

\usepackage{graphicx}
\usepackage{txfonts}
\usepackage{multirow}
\usepackage{colortbl}
\usepackage{multirow}
\usepackage{booktabs}
\usepackage{gensymb}

\usepackage{hyperref}  
\hypersetup{colorlinks=true,linkcolor=[rgb]{1.,0.2,0.2},citecolor=[rgb]{0.1,0.4,1.},filecolor=[rgb]{0.7,0.2,0.2},urlcolor=[rgb]{0.7,0.2,0.2}}

\usepackage{color}
\definecolor{blue}{rgb}{0., 0., 1}
\definecolor{lightblue}{rgb}{0.1,0.4,1.}

\usepackage{orcidlink}

\newcommand {\Tab}{Table\,}
\newcommand {\Sec}{Sect.\,}
\newcommand {\Fig}{Fig.\,}
\newcommand {\Eq}{Eq.\,}

\newcommand {\sn}{$\left<S/N\right>$}

\defcitealias{Granata2025}{G25}

\begin{document}

\title{Enhanced strong-lensing model of MACS~J0138.0$-$2155 based on new JWST and VLT/MUSE observations}
\titlerunning{Enhanced strong-lensing model of MACS~J0138.0$-$2155}

\author{
{A.~Acebron}\inst{\ref{unican}, \ref{inafmilano}} \fnmsep\thanks{E-mail: \href{mailto:ana.acebron@unican.es}{ana.acebron@unican.es}}\orcidlink{0000-0003-3108-9039}
\and
P.~Bergamini\inst{\ref{unimi},\ref{inafbo}}\orcidlink{0000-0003-1383-9414}
\and
P.~Rosati\inst{\ref{unife},\ref{inafbo}}\orcidlink{0000-0002-6813-0632}
\and
P.~Tozzi\inst{\ref{inaffir}}\orcidlink{0000-0003-3096-9966}
\and
M.~Meneghetti\inst{\ref{inafbo}}\orcidlink{0000-0003-1225-7084} 
\and
G.~B.~Caminha\inst{\ref{tum},\ref{mpa}}\orcidlink{0000-0001-6052-3274}
\and
S.~Ertl\inst{\ref{mpa},\ref{tum}}\orcidlink{0000-0002-5085-2143} 
\and
G.~Granata\inst{\ref{unimi},\ref{unife},\ref{icg}}\orcidlink{0000-0002-9512-3788}
\and
{A.~M.~Koekemoer}\inst{\ref{stsci}}\orcidlink{0000-0002-6610-2048}
\and
C.~Grillo\inst{\ref{unimi},\ref{inafmilano}}\orcidlink{0000-0002-5926-7143}
\and
S.~Schuldt\inst{\ref{unimi},\ref{inafmilano}}\orcidlink{0000-0003-2497-6334}
\and
B.~L.~Frye\inst{\ref{arizona}}\orcidlink{0000-0003-1625-8009}
\and
{J.~M.~Diego}\inst{\ref{unican}}\orcidlink{0000-0001-9065-3926}
}

\institute{Instituto de Física de Cantabria (CSIC-UC), Avda. Los Castros s/n, 39005 Santander, Spain \label{unican}
\and
INAF -- IASF Milano, via A. Corti 12, I-20133 Milano, Italy \label{inafmilano}
\and
Dipartimento di Fisica, Universit\`a  degli Studi di Milano, via Celoria 16, I-20133 Milano, Italy \label{unimi}
\and
INAF -- OAS, Osservatorio di Astrofisica e Scienza dello Spazio di Bologna, via Gobetti 93/3, I-40129 Bologna, Italy \label{inafbo}
\and
Dipartimento di Fisica e Scienze della Terra, Universit\`a degli Studi di Ferrara, via Saragat 1, I-44122 Ferrara, Italy \label{unife}
\and
INAF – Osservatorio Astrofisico di Arcetri, Largo E. Fermi 5, I-50125, Firenze, Italy \label{inaffir}
\and
Technical University of Munich, TUM School of Natural Sciences, Physics Department,  James-Franck-Stra{\ss}e 1, 85748 Garching, Germany \label{tum}
\and
Max-Planck-Institut f{\"u}r Astrophysik, Karl-Schwarzschild Stra{\ss}e 1, 85748 Garching, Germany \label{mpa}
\and
Institute of Cosmology and Gravitation, University of Portsmouth, Burnaby Rd, Portsmouth PO1 3FX, UK
\label{icg}
\and
Space Telescope Science Institute, 3700 San Martin Drive, Baltimore, MD 21218, USA \label{stsci}
\and
Department of Astronomy/Steward Observatory, University of Arizona, 933 N Cherry Ave, Tucson, AZ, 85721-0009, USA \label{arizona}
}

   \date{\today}

 
  \abstract
{
We present a new parametric strong-lensing analysis of the galaxy cluster MACS~J0138.0$-$2155 at $z_{\rm L} = 0.336$. This is the first lens cluster known to show two multiply imaged supernova (SN) siblings, SN~Requiem and SN~Encore at $z_{\rm SNe} = 1.949$, enabling a new measurement of the Hubble constant value from the measured relative time delays. 
We exploited Hubble Space Telescope and James Webb  Space Telescope multi-band imaging in synergy with new Multi Unit Spectroscopic Explorer follow-up spectroscopy to develop an improved lens mass model. 
Specifically, we included 84 cluster members ( $\sim60\%$ of which are spectroscopically confirmed) and two perturber galaxies along the line of sight. Our observables consisted of 23 spectroscopically confirmed multiple images from eight background sources, spanning a fairly wide redshift range from 0.767 to 3.420. To accurately characterise the sub-halo mass component, we calibrated the Faber-Jackson scaling relation based on the stellar kinematic measurements of a subset of 14 bright cluster galaxies.
We built several lens models by implementing different cluster total mass parametrisations
to assess the statistical and systematic uncertainties on the predicted values of the position and magnification of the observed and future multiple images of SN~Requiem and SN~Encore. 
Our reference best-fit lens model reproduces the observed positions of the multiple images with a root mean square offset of $0\arcsec.36$ and the multiple-image positions of the SNe and their host galaxy with a remarkable mean precision of only $0\arcsec.05$.
We measure a projected total mass of $M(<60 ~\rm{kpc}) = 2.89_{-0.03}^{+0.04} \times 10^{13} M_{\odot}$, which is consistent with that independently derived from the X-ray analysis of archival data from the Chandra observatory. We also demonstrate the reliability of the new lens model by reconstructing the extended surface-brightness distribution of the multiple images of the host galaxy. The significant discrepancy between the magnification values predicted by our reference model and those from previous studies, which is critical for understanding the intrinsic properties of the two SNe and their host galaxy, further underscores the need to combine cutting-edge observations with a detailed lens modelling.}


   \maketitle
%

\section{Introduction}

While the Hubble Space Telescope (HST) has revolutionised the field of strong lensing in galaxy clusters with its multi-band high-resolution imaging, the James Webb Space Telescope (JWST) has already achieved a similar degree of advancement only three years after its launch. 
In the cluster strong-lensing field, the unprecedented capabilities of the JWST have enabled the identification of large samples of multiple images up to $z~\sim~10$ \citep{Roberts-Borsani2023, Williams2023}. In combination with follow-up spectroscopy from the Multi Unit Spectroscopic Explorer \citep[MUSE,][]{Bacon2010, Bacon2014}, which is mounted on the Very Large Telescope (VLT), a new generation of cluster-scale strong-lensing models have emerged \citep[e.g.][]{Caminha2022b, Bergamini2023b, Diego2024, Furtak2023, Rihtarsic2025}.

Two recent and particularly interesting findings by the JWST have been the discoveries of two supernovae (SNe) that are strongly lensed by galaxy clusters, SN~H0pe \citep{Frye2024, Pierel2024a, Pascale2025} and SN~Encore \citep{Pierel2024b}.
Leveraging the measured time delays between the multiple images of an SN together with a robust total mass reconstruction of the lens galaxy cluster allows for an independent one-step measurement of the Hubble constant ($H_0$) value and the parameter values of the matter density, dark energy density, and dark energy density equation of state \citep[][]{Grillo2024}.

MACS~J0138.0$-$2155 (MACS~J0138, hereafter) is a massive lens galaxy cluster at $z_{\rm L}=0.336$ and the first galaxy cluster that is known to strongly lens two SNe in the same host galaxy at $z=1.949$, called SN~Requiem \citep{Rodney2021} and SN~Encore \citep{Pierel2024b}.
\citet{Rodney2021} reported the discovery of three multiple images of SN~Requiem in archival HST observations of MACS~J0138, which was photometrically classified as a Type Ia SN (SN Ia). They also measured the relative time delays between the multiple images, with significantly large uncertainties ( $\sim 26\%$ for the longest time delay). Three multiple images of SN~Encore were then discovered in the JWST imaging of MACS~J0138, one of which was spectroscopically confirmed as an SN Ia \citep{Pierel2024b, Dhawan2024}. The time delays between the observed multiple images of SN~Encore are measured by Pierel et al. (in prep.).
The study of the strongly-lensed host galaxy, MRG-M0138, is also worth on its own, being is one of the brightest lensed galaxies known in the near-infrared \citep[F160W=17.1,][]{Newman2018a, Newman2018b, Akhshik2020, Akhshik2023, Newman2025}.

An event such as SN Encore provides a unique opportunity to blindly test the validity and robustness of the cluster-scale lens models of MACS~J0138, as previously carried out with SN~Refsdal and SN~H0pe, which are strongly lensed by MACS~J1149.5$+$2223 \citep{Kelly2016, Treu2016, Kelly2023} and PLCK G165.7$+$67.0 \citep{Pascale2025}, respectively. Seven independent groups used different strong-lensing modelling approaches and software and participated in the blind-modelling challenge of the galaxy cluster MACS~J0138, in which the measured time delays between the multiple images of SN~Encore were unknown to the lens modellers. The seven groups agreed on the secure samples of multiple images, cluster members, and line-of-sight perturbers to be included in their independent strong-lensing analyses based on recent JWST and VLT/MUSE observations. The collaboration and the comparison between the seven model predictions of the time-delay and magnification values for both SN~Encore and SN~Requiem are presented by Suyu et al. (in prep.).

In this context, we present here a new strong-lensing model of the galaxy cluster MACS~J0138 with the parametric modelling software \texttt{Lenstool} \citep{Kneib1996, Jullo2007}. This is one of the seven blind lens models submitted to the comparison challenge (see Suyu et al. in prep.). In this paper, we describe our modelling method and forecasts, and we assess the systematic uncertainties.
This paper is organised as follows. In \Sec \ref{sec:data} we describe the HST and JWST imaging, the VLT/MUSE spectroscopic observations, and the archival X-ray observations. Section \ref{sec:SLmodelling} details the method we adopted for the strong-lensing modelling of MACS~J0138. Our results are presented and discussed in \Sec \ref{sec:results}. Finally, our conclusions are summarised in \Sec \ref{sec:conclusions}. 

Throughout the paper, we adopt the standard flat ${\Lambda\mathrm{CDM}}$ cosmological model with $H_0 = 70$ $ \mathrm{km~s^{-1}~Mpc^{-1}}$, $\mathrm{\Omega_{m}}=0.3$, and $\mathrm{\Omega_{\Lambda}}=0.7$. In this cosmology, $1\arcsec$ corresponds to a physical scale of 4.81 kpc at the lens cluster redshift ($z_{\rm L}=0.336$).
The magnitudes are quoted in the AB system \citep{Oke1974}. The statistical uncertainties are given as the 68\% confidence interval.

\section{Multi-wavelength observations of MACS~J0138.0$-$2155} \label{sec:data}
This section briefly reviews the HST and JWST multi-band imaging, the VLT/MUSE spectroscopy, and the X-ray data sets we used. We refer to \citet{Pierel2024b} and \citet[][\citetalias{Granata2025} hereafter]{Granata2025} for a detailed overview.

MACS~J0138 has been repeatedly surveyed by the HST and the JWST. Specifically, the lens cluster has been imaged within three HST programmes (P.I.: Newman, P.I.: Akhshik, and P.I.: Pierel) that collected observations with eight HST bands (F390W, F555W, F814W, F105W, F125W, F140W, and F160W). Following the discovery of SN Encore, the galaxy cluster was imaged with six JWST bands (F115W, F150W, F200W, F277W, F356W, and F444W) within two JWST programmes (P.I.: Newman, and P.I.: Pierel).
All the HST and JWST imaging data were astrometrically aligned relative to each other and to the Gaia-DR3 absolute astrometric frame, and they were combined into mosaics with a pixel size of $0.\arcsec 04$, following the approaches first described by \citet{Koekemoer2011} and updated as needed \citep[see][for more details]{Pierel2024b}. A summary of the HST and JWST observations is given in Tables 1 and 2 in \citet{Ertl2025}.
The photometry of all the galaxies \citep[with \texttt{MAG\_AUTO} $<23$ in the F160W band, as measured from \texttt{Source-Extractor,}][]{Bertin1996} in the field of view is presented by \citet{Ertl2025}. The surface-brightness distribution of the cluster members was modelled with the software \texttt{Morphofit} \citep{Tortorelli2023a}, which has proven to work well in highly crowded fields \citep{Tortorelli2018, Tortorelli2023b}. Within an automated and parallelised framework, \texttt{Morphofit} uses both \texttt{Source-Extractor} and \texttt{Galfit} \citep{Peng2011} to estimate the galaxies structural parameters by fitting multiple surface-brightness components. \citet{Ertl2025} used several Sérsic profiles (when needed) to accurately estimate the multi-band photometry of 81 cluster member galaxies (i.e. for all but the three jellyfish galaxies).

For the spectroscopy, MACS~J0138 was targeted by two VLT/MUSE programmes (IDs 0103.A-0777, P.I.: Edge and 110.23PS.016, P.I.: Suyu), which amounted to a cumulative exposure time of about 3.7 hours. The observations, data reduction, redshift, and stellar kinematic measurements are described in \citetalias{Granata2025}. We used the spectroscopically confirmed multiple images and cluster member galaxies to build our new strong-lensing model (see \Sec \ref{sec:SLmodelling} for further details). 

MACS~J0138 was also observed by the Chandra telescope on 2015 June 11-12 with the Advanced CCD Imaging Spectrometer (ACIS-I; Observation ID 17186; P.I.: Boehringer) for a total exposure time of 24.7 ks after the data reduction. 
The data reduction was performed with the same procedure as was described by \citet{2022aTozzi}, starting from the level 1 event files with {\tt CIAO 4.15} and the latest release of the Chandra Calibration Database {\tt (CALDB 4.10.4)}.  
The VFAINT observation mode allowed us to reject a significant fraction of the ACIS particle background.
A small number of source photons ($46\pm 7$) were filtered out due to pile-up, given the brightness of the cool core. We verified a posteriori that this effect did not impact the spectral analysis of the nucleus, and therefore we kept the standard data reduction in the VFAINT mode to obtain a lower background and achieve a better characterisation of the diffuse intracluster medium (ICM) emission. The images were created in the 0.5-2 keV, 2-7 keV, and total (0.5-7 keV) bands without binning to preserve the full angular resolution of $\sim 1$ arcsec (full width at half maximum) at the aimpoint (1 pixel corresponds to 0.492 arcsec).  
Based on the bright ICM emission of the galaxy cluster, we were able to detect about 7350 net counts within the 0.5-7 keV band. The image quality and the high signal from the ICM allowed us to perform a spatially resolved spectral analysis and, ultimately, to derive the hydrostatic mass profile (see \Sec\,\ref{sec:massdistribution}).

\begin{figure*}
        \centering
        \includegraphics[scale=0.7]{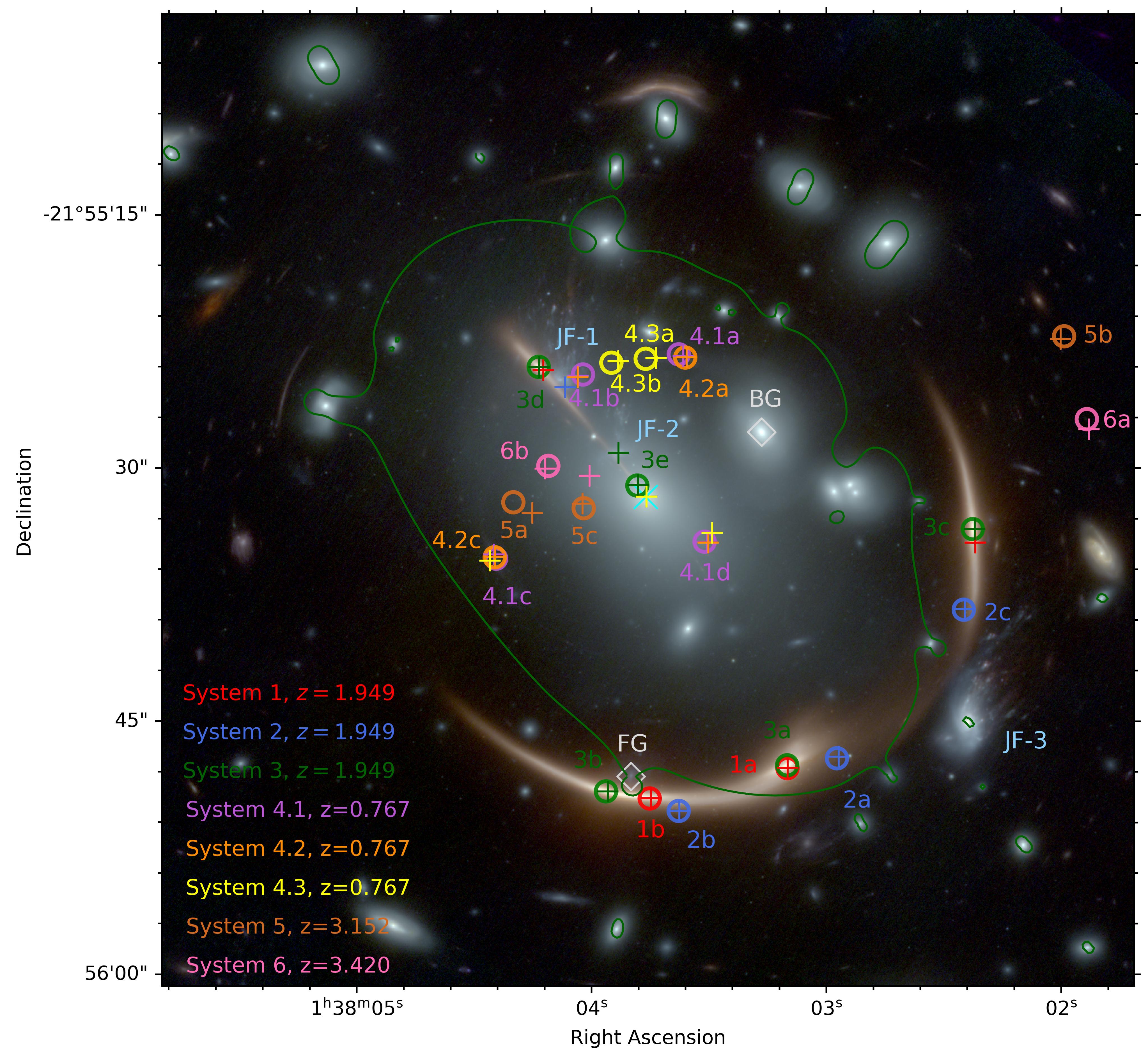}
        \caption{JWST colour-composite image of MACS~J0138$-$2155, created with the passbands F115W$+$F150W (blue), F200W$ + $F277W (green), and F356W$ + $F444W (red). The 23 spectroscopically confirmed multiple images included in our strong-lensing analysis are marked with circles, colour-coded per system, as specified in the legend. The crosses, following the same colour-coding scheme, mark the best-fit model-predicted positions of the multiple images from our \texttt{reference} strong-lensing model. The two line-of-sight galaxies and the three jellyfish member galaxies included in our model are labelled in grey and light blue, respectively, following \Sec \ref{sec:SLmodelling}. The best-fit critical curves from our \texttt{reference} strong-lensing model at the redshift of the SNe host galaxy ($z$ = 1.949), are shown in green. The cyan cross indicates the reference position, associated with the BCG.}
        \label{fig:macs0138}
\end{figure*}

\section{Strong lensing modelling} \label{sec:SLmodelling}
We modelled the total mass distribution of MACS~J0138 with the publicly available and parametric pipeline \texttt{Lenstool}\footnote{\url{https://projets.lam.fr/projects/lenstool}} \citep{Kneib1996, Jullo2007}, which exploits a Bayesian Markov chain Monte Carlo (MCMC) sampler. The lens model we present was developed within the modelling challenge presented by Suyu et al. (in prep.). The seven independent participating teams agreed upon the selection of the observables and cluster members. 
We thus refer to \citet{Ertl2025}, \citetalias{Granata2025}, and Suyu et al. (in prep.) for a detailed overview of the multiple images and cluster member selections and only provide a summary below. 

We considered the gold sample of multiple images described by \citet{Ertl2025} and Suyu et al. (in prep.) as lensing observables. Briefly, these are 23 spectroscopically confirmed multiple images from eight background sources spanning a redshift range $0.767 < z_{\rm s} < 3.420$ 
\citep[see \Tab 5 in][]{Ertl2025}. For clarity, we adopted the same labelling as used by \citet{Ertl2025} and Suyu et al. (in prep.). The gold sample of multiple images considered in our lensing analysis is shown in \Fig \ref{fig:macs0138}. In particular, the two most noteworthy strongly lensed systems are systems 1 and 2, which correspond to the multiple images of SN Encore and SN Requiem, respectively, at $z_{\rm s1, s2} = 1.949$.

The cluster member catalogue includes 84 galaxies with F160W$<24$, of which $\sim 60\%$ are spectroscopically confirmed through the VLT/MUSE observations. The 50 spectroscopic cluster members were selected as the galaxies with rest-frame line-of-sight velocities within $\Delta V = 2500 \rm ~ km ~s^{-1}$ from the galaxy cluster mean line-of-sight velocity (\citetalias{Granata2025}).
The photometric cluster galaxies were firstly selected as those lying within $2\sigma$ of the red cluster sequence (defined in the F555W$-$F814W versus F814W colour-magnitude diagram). This sample was then complemented with a F277W$-$F444W versus F814W$-$F150W colour-colour selection, resulting in the inclusion of 10 additional member galaxies \citep{Ertl2025}.
We remark that three cluster members are jellyfish galaxies, which are labelled JF-1, JF-2, and JF-3 in the following (see also \Fig\,\ref{fig:macs0138}).

\begin{figure}
        \centering
        \includegraphics[scale=0.73]{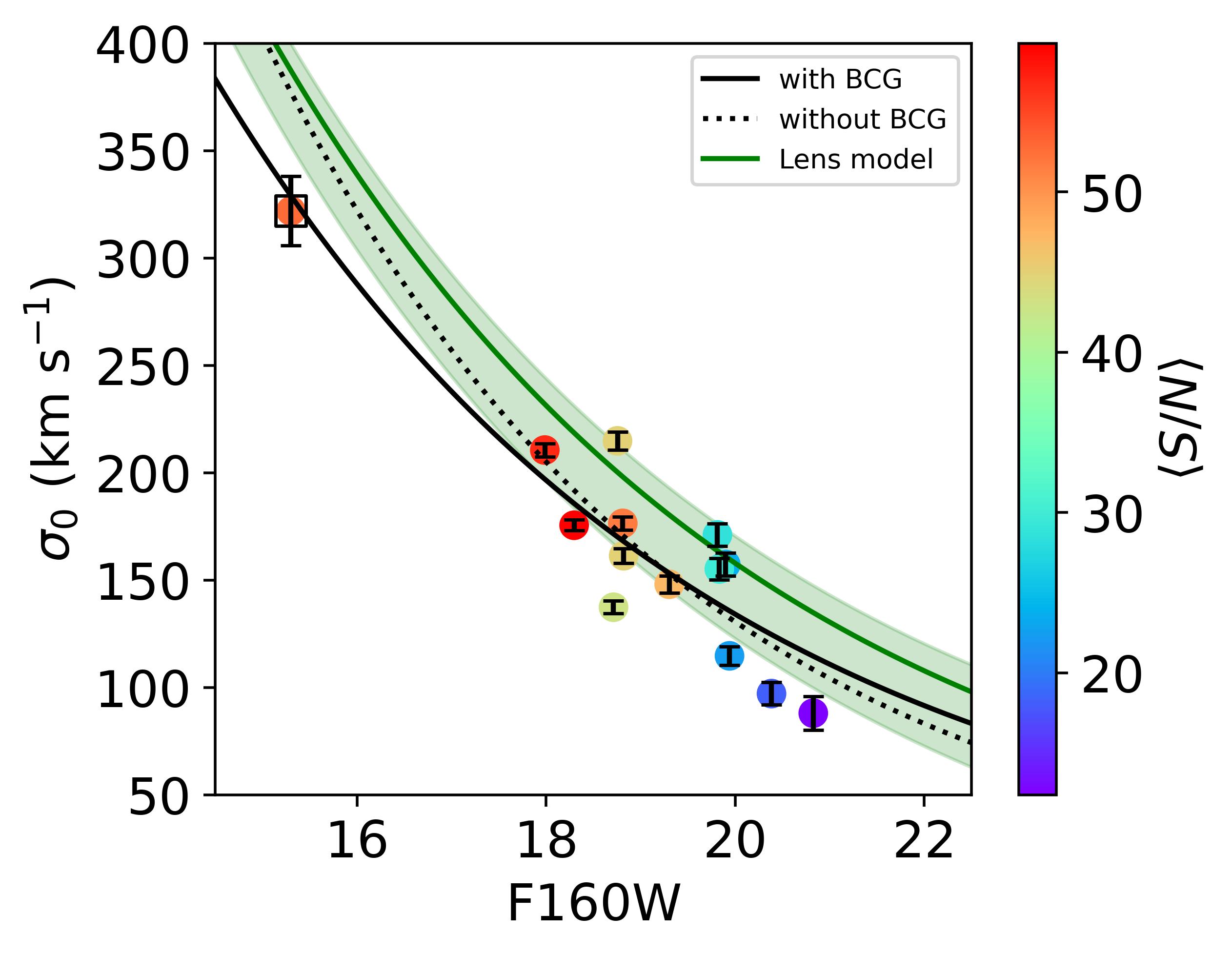}
        \caption{Calibration of the Faber-Jackson relation. The filled circles show the measured stellar velocity dispersions, $\sigma_0$, of a subset of 14 bright cluster member galaxies as a function of their total magnitudes in the HST F160W band (colour-coded depending on their averaged spectral \sn\ values) from \citetalias{Granata2025}. The black square highlights the $\sigma_0$ measurement of the BCG. The solid and dotted black lines correspond to the best-fit $\sigma_0$-F160W relations, obtained as described in \Sec \ref{sec:SLmodelling}. The green line and shaded area show the median and the 68\% confidence level of the $\sigma_0$-F160W relation from our \texttt{reference} strong-lensing model.}
        \label{fig:sigmaSNEncore}
\end{figure}

We further exploited the stellar velocity dispersion measurements for a sub-set of cluster galaxies obtained by \citetalias{Granata2025} to calibrate the galaxy cluster sub-halo mass component. In detail, \citetalias{Granata2025} present a robust stellar kinematic measurement of 14 bright cluster member galaxies, with high averaged spectral signal-to-noise ratios (\sn$~ \geq 10$), down to F160W $\sim 21$ (see \Fig\,\ref{fig:sigmaSNEncore}).
Following \citet[][see their Appendix B]{Bergamini2019}, we adopted a Bayesian approach by making use of the Python implementation of the affine-invariant MCMC ensemble sampler \citep{Goodman-Weare2010, Foreman-Mackey2013}. This allowed us to sample the posterior distribution and estimate the best-fit values of the reference central stellar velocity dispersion, $\sigma_0^{\rm ref}$, its scatter, $\Delta \sigma_0^{\rm ref}$, and the logarithmic slope, $\alpha$, of the $\sigma_0-$F160W relation (see \Sec \ref{sec:SLmethod}). 
As shown in \Fig\,\ref{fig:sigmaSNEncore}, the brightest cluster galaxy (BCG) seems to be an outlier in this relation, possibly due to its different assembly history compared to that of other galaxies. We thus performed two fits of the $\sigma_0-$F160W relation, with and without the BCG. We present the results of the fits in \Tab\,\ref{tab:fit_stellarkinematics}, that is, the best-fit, together with the median and the 16th and 84th percentile values of $\sigma_0^{\rm ref}$, $\Delta \sigma_0^{\rm ref}$, and $\alpha$, as obtained from their posterior probability distributions.
These values were then considered as prior information for the scaling relations of the cluster members in \Sec\,\ref{sec:models}. 

We note that the values of the fitted relation without the BCG are perfectly consistent with those derived by \citetalias{Granata2025}, as both analyses implemented the same method. Our estimates also agree with those presented by \citet{Flowers2024}, who exploited shallower VLT/MUSE data (associated with the programme 0103.A-0777) and applied different quality cuts from those by \citetalias{Granata2025}.

\subsection{Modelling method} \label{sec:SLmethod}
In a parametric lens model of MACS~J0138, its total mass distribution can be separated into several mass components. We modelled all haloes with dual pseudo-isothermal elliptical (dPIE) mass density profiles \citep{Eliasdottir2007, Suyu2010}, characterised by seven free parameters in \texttt{Lenstool}. These are the centre coordinates, $x,~y$;  the ellipticity, $e=(a^2-b^2)/(a^2+b^2)$, where $a$ and $b$ are the values of the major and minor semi-axes, respectively; the orientation, $\theta$ (counted counterclockwise from the $x$-axis); the core and truncation radii, $r_{\rm core}$ and $r_{\rm cut}$, respectively; and a velocity dispersion, $\sigma_{\rm LT}$, which is related to the central velocity dispersion of the dPIE mass density profile through $\sigma_0 = \sqrt{3/2}~ \sigma_{\rm LT}$.

\begin{table*}
  \renewcommand\arraystretch{1.2}

  \setlength\tabcolsep{0.4em}
  \centering
  \caption{Results from the fit of the stellar kinematic measurements from \citetalias{Granata2025}, as described in \Sec~\ref{sec:SLmodelling}.
}
  \begin{tabular}{ccccc}
    \hline
    \hline
    &\multicolumn{2}{c}{with the BCG} &
    \multicolumn{2}{c}{without the BCG} \\
    \cmidrule(r){2-3}\cmidrule(l){4-5}
    Parameter & Best-fit value & Marginalised value & Best-fit value & Marginalised value\\
    \hline
    $\sigma_0^{\rm ref}$ [$\rm km~s^{-1}$] & $329$ & $329^{+29}_{-28}$ &  $377$ & $394^{+95}_{-80}$\\
    $\Delta \sigma_0^{\rm ref}$ [$\rm km~s^{-1}$] & $26$ & $26^{+7}_{-5}$ &  $23$ & $27^{+7}_{-5}$\\
    $\alpha$ & $0.21$ & $0.21^{+0.03}_{-0.03}$ &  $0.25$ & $0.26^{+0.06}_{-0.06}$\\
    \hline
  \end{tabular}
\tablefoot{
Best-fit and marginalised parameter values with the associated 1$\sigma$ uncertainties for a reference galaxy, assumed to be the BCG, with a total magnitude value F160W$=15.30$.
}
   \label{tab:fit_stellarkinematics}
   \end{table*}

We described the cluster-scale halo with a non-truncated, elliptical dPIE mass density profile, while we considered singular circular dPIE profiles for the galaxy-scale haloes, which are associated with the 84 cluster members and the two line-of-sight galaxies. 
To significantly reduce the number of free parameters, the total masses of the early-type galaxy-scale haloes were scaled with total mass-to-light ratios increasing with their near-IR luminosities \citep[which are good proxy of their stellar mass,][]{Grillo2015}. More specifically, we adopted the following two scaling relations \citep{Jullo2007}:

\begin{equation}
\label{eqSR}
\sigma_0=\sigma_0^{\rm ref} \left(\frac{L}{L^{\rm ref}}\right)^{\alpha}
\text{and}~\,
r_{\rm cut}=r_{\rm cut}^{\rm ref}\left(\frac{L}{L^{\rm ref}}\right)^{\beta},
\end{equation}

where $L^{\rm ref}$ represents the reference luminosity value of a galaxy at the cluster redshift, which we associated with the BCG (with a total magnitude value in the HST F160W band of $15.30$). The parameters $\alpha$ and $\beta$, corresponding to the slopes of the $\sigma_0$ and $r_{\rm cut}$ scaling relations, respectively, were fixed in the lens model, and thus, the sub-halo mass component is completely described with only two free parameters: $\sigma_0^{\rm ref}$ and $r_{\rm cut}^{\rm ref}$.
In our lens models, we used the normalisation and slope values of the best-fit $\sigma_0$--F160W relations (see \Fig \ref{fig:sigmaSNEncore} and \Tab \ref{tab:fit_stellarkinematics}), whilst a large flat prior for the $r_{\rm cut}^{\rm ref}$ value was considered. 
In turn, the adopted value of $\beta$ was inferred, following $\beta =\gamma-2\alpha+1$, where $\gamma$ represents the slope of the mass-to-light ratio. We adopted $\gamma=0.2$, entailing that the galaxy total mass-to-light ratio varies with the luminosity as $M^{\rm tot}_i L_i^{-1} \propto L ^{0.2}_i$, which is consistent with the fundamental plane \citep{Faber1987, Bender1992}.

Finally, we introduced further flexibility into the lens modelling of MACS~J0138 by including an external shear component, characterised by two additional free parameters: its magnitude, $\gamma_{\rm ext}$, and orientation, $\phi_{\rm ext}$. Previous studies have supported the use of this component to significantly improve the reproduction of the observed multiple image positions by limiting the effects of unmodelled mass structures present in the lens environment or along the line of sight, or of the complex internal structure of the lens \citep[e.g.][]{Lagattuta2019, Acebron2022, RafeeAdnan2024}, as further discussed in \Sec~\ref{ENV}.

The best-fit values of the model parameters that describe the total mass distribution of MACSJ 0138, $\mathbf{p}$, were obtained by minimising on the image plane the distance between the observed, $\boldsymbol{\theta}^{\rm obs}$, and model-predicted, $\boldsymbol{\theta}^{\rm pred}$, positions of the multiple images through the following $\chi^2$ function:
\begin{equation}
\chi^2(\mathbf{p})=\sum_{j=1}^{N_{\rm fam}}\sum_{i=1}^{N_{\rm img}^{j}}\left(\frac{\left|\boldsymbol{\theta}^{\rm obs}_{ij}-{\boldsymbol{\theta}}^{\rm pred}_{ij}(\mathbf{p})\right|}{\sigma_{ij}}\right)^2, 
\end{equation}
where $N_{\rm fam}$ and $N_{\rm img}^{j}$ are the total number of families and of multiple images for the family $j$ included in the lens model, respectively, and $\sigma_{ij}$, the positional uncertainty for each multiple image.
We assumed initial circular values of $\sigma_{ij}$ based on the observed spatial extent of the multiple images. For each multiple image, the circularised uncertainty was obtained as $\sigma=\sqrt{\sigma_a \sigma_b}$, where $\sigma_a$ and $\sigma_b$ are the semi-major and semi-minor axis of the error ellipse, respectively, as given in \Tab 5 in \citet{Ertl2025}. For each final model optimisation, the multiple image positional uncertainty was rescaled to obtain a minimum $\chi^2$ value comparable with the number of degrees of freedom ($\nu$), or $\chi^2/\nu\sim1$. This ensures that the posterior distributions of the model parameters are properly sampled to estimate robust and meaningful uncertainties. The median values and the associated $1\sigma$ uncertainties of the model parameters were instead extracted from the MCMC chains, which have a total number of $5\times 10^5$ samples, excluding the burn-in phase. 

We also used the root mean square (rms) value of the difference between the observed and model-predicted positions of the multiple images to quantify the goodness of our models, 
\begin{equation}
{\rm rms}=\sqrt{\frac{1}{N_{\rm tot}}\sum_{i=1}^{N_{\rm tot}}\left|\boldsymbol{\theta}^{\rm obs}_{i}-{\boldsymbol{\theta}}^{\rm pred}_{i}(\mathbf{p})\right|^2},
\end{equation}
where $N_{\rm tot}$ is the total number of multiple images.
To robustly compare lens models with different numbers of free parameters, however, we also employed other statistical estimators, such as the logarithm of the maximum value of the likelihood, $\mathcal{L}$; and of the evidence, $E$; the Bayesian information criterion \citep[BIC;][]{Schwarz1978}; and the Akaike information criterion \citep[AIC;][]{Akaike1974}. These are defined as follows:
\begin{equation}
{\rm BIC}=-2\ln({\mathcal{L}}) + N_{\rm param} \times \ln(n),
\end{equation}
and 
\begin{equation}
{\rm AIC}=2N_{\rm param} -2\ln({\mathcal{L}}),
\end{equation}
where $N_{\rm param}$ and $n$ are the number of free parameters and observables, respectively. 

\subsection{Total mass models} \label{sec:models}
Investigating possible systematic uncertainties, arising from lens modelling choices of the total mass reconstruction of MACS~J0138, is important given the prospects of using this lens cluster for time-delay cosmography.
To do this, we explored several total mass parametrisations of the cluster and assessed the robustness of the model-predicted positions and magnifications of the SN~Requiem and SN~Encore multiple images. To ensure a blind cosmological analysis \citep[eliminating any experimenter bias, see e.g.][]{Pascale2025} both in the lens modelling and in the time-delay measurement analyses (Pierel et al. in prep.), the model-predicted time delays between the SNe multiple images and their uncertainties are not presented in this publication, but will instead be discussed by Suyu et al. (in prep.).

All of our lens models of MACS~J0138 considered a single main cluster-scale halo, parametrised with a non-truncated dPIE mass density profile (see \Sec~\ref{sec:SLmodelling}). All models but one included an external shear component. The sub-halo component and the perturber galaxies, such as the jellyfish members or the line-of-sight galaxies, were parametrised with singular circular dPIE mass density profiles. The three jellyfish member galaxies were modelled outside of the scaling relations. We also included the mass contribution from two line-of-sight perturber galaxies (see \citetalias{Granata2025}), the foreground spiral FG ($z_{\rm FG}=0.309$) and the background BG ($z_{\rm BG}= 0.371$). Their positions are highlighted in \Fig~\ref{fig:macs0138}. The two line-of-sight galaxies were modelled at the lens cluster redshift ($z_{\rm L}= 0.336$). As shown in \citet{Acebron2022}, when the line-of-sight perturbers are close to the redshift of the main deflector (the redshift difference of both FG and BG with respect to MACS~J0138 is $\leq 10\%$), this approximation limits biases on the model-predicted magnifications and time delays between multiple images that are angularly close to the perturbers (as is the case for FG, with systems 1, 2, and 3, see \Fig~\ref{fig:macs0138}), compared to models that do not include such perturbers.
The values of all the model parameters, but that of the scaling relation normalisation, $\sigma_{0}^{\rm ref}$, were optimised within large flat priors, as shown in \Tab~\ref{table:inout_lensing} (top). 
Below, we detail the specific characteristics of each lens model we built. 
The following four lens models were created during the blind comparison challenge described by Suyu et al. (in prep.).

\begin{itemize}
\item \textbf{\texttt{reference}}:
By leveraging the calibrated Faber-Jackson relation including the BCG (see \Tab~\ref{tab:fit_stellarkinematics} and the solid line in \Fig~\ref{fig:sigmaSNEncore}), the sub-halo mass component that is associated with the 81 cluster member galaxies, including the BCG, was modelled within the scaling relations, as introduced in \Sec~\ref{sec:SLmethod}.
We thus adopted a Gaussian distribution centred on the measured value of $\sigma_{0}^{\rm ref}$, with a standard deviation value equal to $\Delta \sigma_{0}^{\rm ref}$, as a prior for the normalisation of the kinematic scaling relation. This reference strong-lensing model was submitted for the modelling challenge presented by Suyu et al. (in prep.).\\

\item \textbf{\texttt{circularBCGout}}: We instead exploited the calibrated Faber-Jackson relation without the BCG (see \Tab~\ref{tab:fit_stellarkinematics} and the dotted line in \Fig~\ref{fig:sigmaSNEncore}) to scale the mass-to-light ratio of 80 cluster member galaxies. The BCG was modelled outside of the scaling relations, with a circular dPIE mass density profile, where the values of its central stellar velocity dispersion and truncation radius were optimised within large flat priors. \\

\item \textbf{\texttt{ellipticalBCGout}}:
In this model, the BCG was instead modelled with an elliptical dPIE mass density profile, where the values of the ellipticity and position angle were also freely optimised  within large flat priors. \\

\item \textbf{\texttt{reference-noFG}}: The foreground spiral galaxy FG lies at a projected distance of only $\sim1\arcsec.7 ~ (3\arcsec.4)$ from the key multiple image systems 1 (SN~Encore) and 3 (2, SN~Requiem). To explore the impact of this perturber on our results, we considered the \texttt{reference} lens model, but without the mass contribution associated with FG.
\end{itemize}

The last two models were instead developed after the unblinding with the other lens modelling teams. We remark, however, that, importantly, the time delays between the SNe multiple images are still unknown at the time of writing.
 
\begin{itemize}
\item \textbf{\texttt{reference-subs}}: Interestingly, wide-field imaging from the DESI Legacy Imaging Surveys of MACS~J0138 reveals a massive structure $140\arcsec$ or $673~\rm kpc$ south-east of the BCG. This massive substructure is located at the edge of the JWST imaging of the MACS~J0138 in the F150W and F444W bands (see \Sec~\ref{ENV}). The galaxies colours and photometric redshifts are consistent with those of the spectroscopically confirmed cluster members in the main cluster core. We included this massive structure as a singular isothermal sphere (SIS) mass density profile. Its position was fixed to that of the brightest galaxy at (RA, Dec)=($24.525567\degree$, $-21.964201\degree$), and its velocity dispersion value was optimised within a flat large prior, that is, we added one more free parameter.\\

\item \textbf{\texttt{subs-noES}}: The last lens model was the same as the previous model, without an external shear component. Its scope was to test whether the external shear might be ascribed to the non-modelled lens cluster environment alone.
\end{itemize}

The number of free model parameters ($N_{\rm param}$) and degrees of freedom ($\nu$) characterising each of the lens models explored are given in \Tab\,\ref{tab:model_stat}.

\begin{figure*}
        \centering
        \includegraphics[scale=0.64]{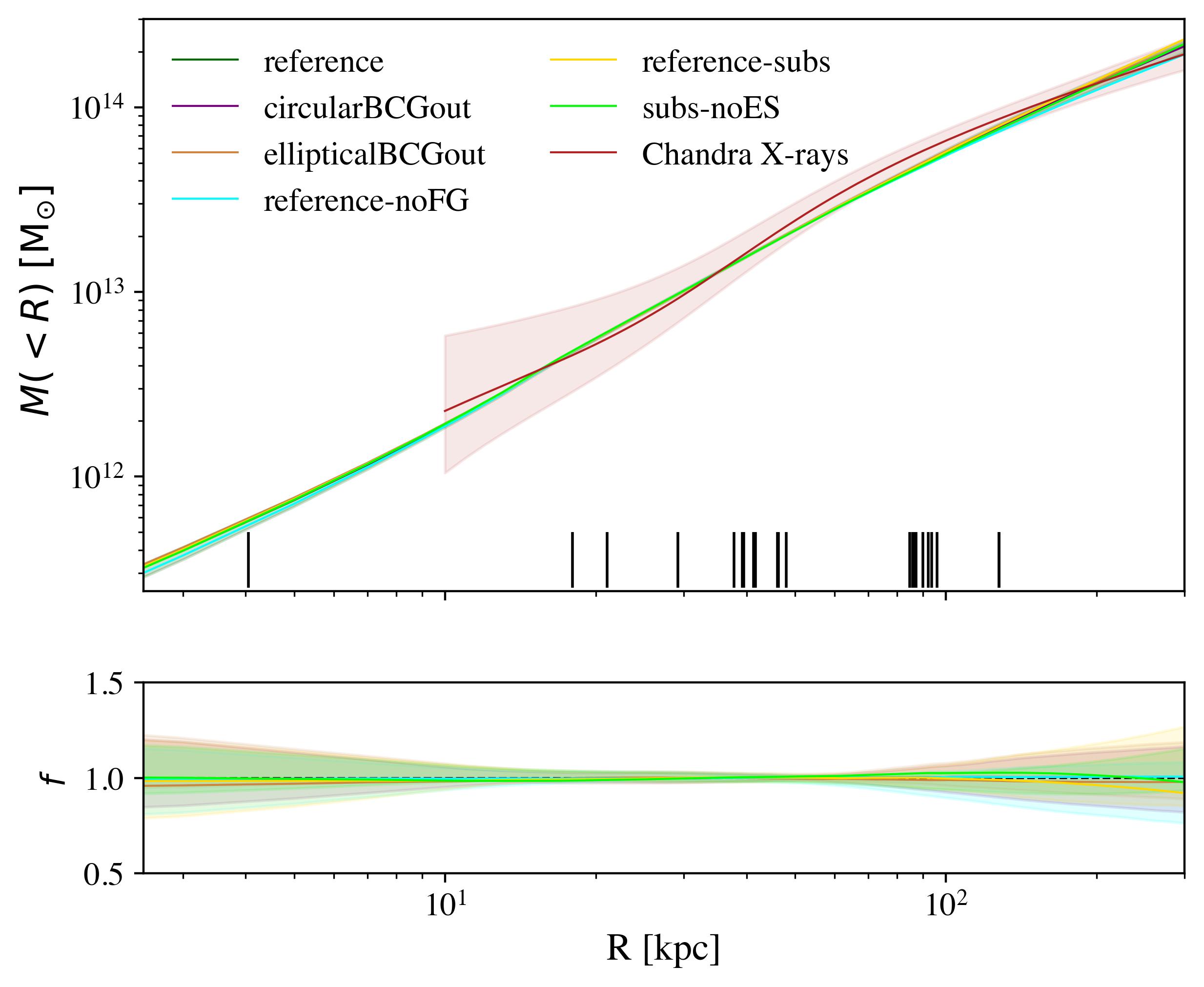}
        \includegraphics[width=0.91\columnwidth]{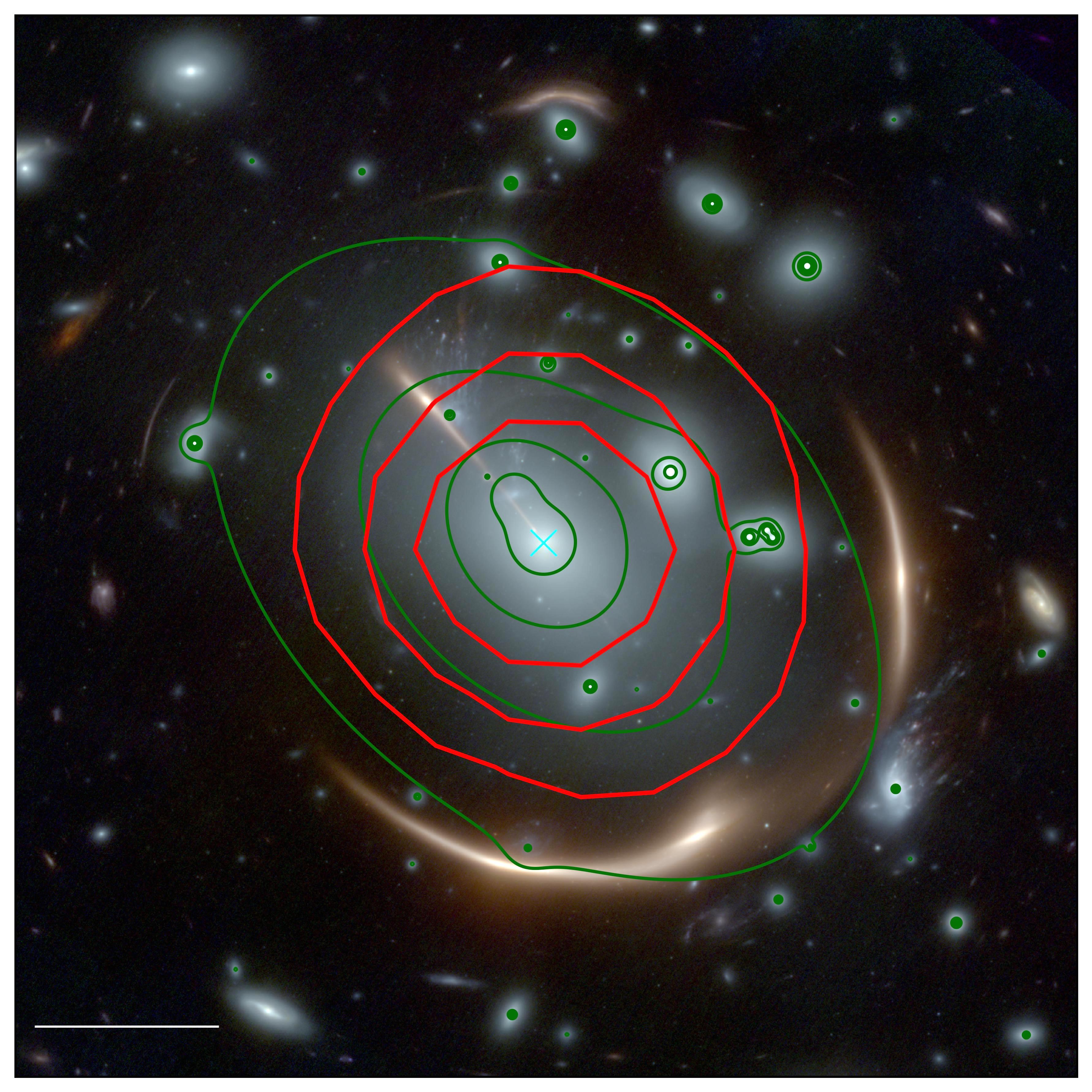}
        \caption{Total mass distribution of MACS~J0138. \textbf{Left:} Cumulative projected total mass profile of the lens cluster as a function of the distance from the BCG (highlighted with a cyan cross in \Fig\,\ref{fig:macs0138}). The coloured lines show the median values of the total mass profile, and the shaded regions encompass the 16th and the 84th percentiles, estimated from 1000 random MCMC realisations. The vertical black lines denote the projected distances of the 23 multiple images from the BCG. The bottom panel illustrates the ratio of the projected total mass profiles for the different models and the best-fitting \texttt{reference} one. The cumulative projected hydrostatic mass profile measured from the analysis of the Chandra data, with the associated $1 \sigma$ uncertainties, is shown in dark red.
        \textbf{Right:} Colour-composite JWST image of MACS~J0138, where we overlay the contour levels ($[1.5, 2.2, 3.0, 4.5] \times 10^9 M_{\odot}~ \rm kpc^{-2}$) of the total projected surface mass density distribution from our best-fit \texttt{reference} strong-lensing model (in dark green) and the Chandra X-ray surface brightness (in red). The white bar represents a scale of $10\arcsec$. North is up, and east is left.}
        \label{fig:massdist_SNEncore}
\end{figure*}

\begin{table*}[ht!]   
    \tiny
	\def\arraystretch{2.}
    \setlength\tabcolsep{0.4em}
	\centering          
  \caption{Input and output parameters of the \texttt{reference} lens model for the galaxy cluster MACS~J0138. The $x$ and $y$ coordinates are expressed in arcseconds with respect to the position of the BCG, at RA=24.51570318, Dec=$-$21.92547911. The values in square brackets are kept fixed during the model optimisation, while those that are separated by the $\div$ symbol correspond to the lower and upper boundaries of the flat prior assumed in the model. Instead, the symbol $\pm$ specifies a Gaussian prior where we quote the distribution's mean and standard deviation. We quote, in the bottom table, the median and the 16th, and 84th percentiles from the model marginalised posterior distributions of the 20 model free parameters.}
	\begin{tabular}{|c|c|c|c|c|c|c|c|c|c|}
	    \cline{2-10}
		\multicolumn{1}{c|}{} & \multicolumn{9}{c|}{ \textbf{Input parameter values and assumed priors}} \\
		\cline{2-10}
		  \multicolumn{1}{c|}{} & \boldmath{$x\, \mathrm{[\arcsec]}$} & \boldmath{$y\, \mathrm{[\arcsec]}$} & \boldmath{$e$} & \boldmath{$\theta\ [^{\circ}]$} & \boldmath{$\sigma_{\rm LT}\, \mathrm{[km\ s^{-1}]}$} & \boldmath{$r_{\rm core}\, \mathrm{[\arcsec]}$} & \boldmath{$r_{\rm cut}\, \mathrm{[\arcsec]}$} & \boldmath{$\gamma_{\rm ext}$} & \boldmath{$\phi_{\rm ext}\ [^{\circ}]$} \\ 
          \hline
		  
		   \boldmath{\bf{Cluster-scale halo}} & $-7.0\,\div\,7.0$ & $-7.0\,\div\,7.0$ & $0.0\,\div\,0.9$ & $0.0\,\div\,180.0$ & $400\,\div\,1500$ & $0.0\,\div\,20.0$ & $[2000.0]$ & -- & -- \\

\boldmath{\bf{External shear}} & -- &  --  &  --  &  --  &  --  &  -- & -- & $0.0\,\div\,1.0$ & $0.0\,\div\,180.0$ \\
          
          \hline
          \multicolumn{1}{c}{}
          \\[-5ex]

          \hline
		  
		   \bf{JF-1} & $[-5.11]$ & $[6.94]$ & $[0.0]$ & $[0.0]$ & $40\,\div\,250$ & $[0.001]$ & $0.0\,\div\,30.0$ & -- & -- \\
		  
  		   \bf{JF-2} & $[-1.65]$ & $[2.60]$ & $[0.0]$ & $[0.0]$ & $40\,\div\,250$ & $[0.001]$ & $0.0\,\div\,30.0$ & -- & -- \\

          \bf{JF-3} & $[19.09]$ & $[-13.36]$ & $[0.0]$ & $[0.0]$ & $40\,\div\,250$ & $[0.001]$ & $0.0\,\div\,30.0$ & -- & -- \\

           \bf{BG} & $[6.87]$ & $[3.84]$ & $[0.0]$ & $[0.0]$ & $100\,\div\,300$ & $[0.001]$ & $0.0\,\div\,30.0$ & -- & -- \\

           \bf{FG} & $[-0.87]$ & $[-16.56]$ & $[0.0]$ & $[0.0]$ & $20\,\div\,250$ & $[0.001]$ & $0.0\,\div\,30.0$ & -- & -- \\
		  
		  \cline{1-10}
                            
		   \bf{Scaling relations} & $\boldsymbol{N_{\rm gal}=}81$
		  & $\boldsymbol{{\mathrm{F160W}}^{\rm ref}=}15.30$
		  & $\boldsymbol{\alpha=}0.21$ & $\boldsymbol{\sigma_{\rm LT}^{\rm ref}=}268\,\pm\,30$ & $\boldsymbol{\beta=}0.78$ & $\boldsymbol{r_{\rm cut}^{\rm ref}=}1.0\,\div\,100.0$ & $\boldsymbol{\gamma=}0.20$ & -- & --\\

		  \hline
		 
	\end{tabular}
	\\[2ex]
    
	\begin{tabular}{|c|c|c|c|c|c|c|c|c|c|}
	    \cline{2-10}
		\multicolumn{1}{c|}{} & \multicolumn{9}{c|}{ \textbf{Optimised parameter values}} \\
		\cline{2-10}
		  \multicolumn{1}{c|}{} & \boldmath{$x\, \mathrm{[\arcsec]}$} & \boldmath{$y\, \mathrm{[\arcsec]}$} & \boldmath{$e$} & \boldmath{$\theta\ [^{\circ}]$} & \boldmath{$\sigma_{\rm LT}\, \mathrm{[km\ s^{-1}]}$} & \boldmath{$r_{\rm core}\, \mathrm{[\arcsec]}$} & \boldmath{$r_{\rm cut}\, \mathrm{[\arcsec]}$} & \boldmath{$\gamma_{\rm ext}$} & \boldmath{$\phi_{\rm ext}\ [^{\circ}]$} \\ 
          \hline
  
		   \boldmath{\bf{Cluster-scale halo}} & $0.44^{+0.52}_{-0.45}$ & $-1.59^{+0.63}_{-0.90}$ & $0.39^{+0.03}_{-0.04}$ & $138.5^{+1.8}_{-1.5}$ & $879^{+35}_{-59}$ & $9.9^{+1.0}_{-1.7}$ & $[2000.0]$  & -- & --\\
          
 \boldmath{\bf{External shear}} & -- &  --  &  --  &  --  &  --  &  -- & -- & $0.11^{+0.02}_{-0.02}$ & $155.6^{+7.7}_{-5.8}$ \\

          \hline
          \multicolumn{1}{c}{}
          \\[-5ex]

          \hline
		  
		 \bf{JF-1} & $[-5.11]$ & $[6.94]$ & $[0.0]$ & $[0.0]$ & $69^{+33}_{-19}$ & $[0.001]$ & $12.4^{+12.0}_{-8.9}$  & -- & --\\
		  
  		 \bf{JF-2} & $[-1.65]$ & $[2.60]$ & $[0.0]$ & $[0.0]$ & $243^{+5}_{-14}$ & $[0.001]$ & $6.3^{+7.3}_{-2.1}$  & -- & --\\

         \bf{JF-3} & $[19.09]$ & $[-13.36]$ & $[0.0]$ & $[0.0]$ & $56^{+53}_{-13}$ & $[0.001]$ & $0.8^{+6.1}_{-0.7}$  & -- & --\\

         \bf{BG} & $[6.87]$ & $[3.84]$ & $[0.0]$ & $[0.0]$ & $167^{+29}_{-20}$ & $[0.001]$ & $20.4^{+6.9}_{-12.1}$ & -- & --\\

         \bf{FG} & $[-0.87]$ & $[-16.56]$ & $[0.0]$ & $[0.0]$ & $54^{+23}_{-7}$& $[0.001]$ & $7.1^{+14.2}_{-6.2}$ & -- & --\\
		  
		  \cline{1-10}
		  
		   \bf{Scaling relations} & $\boldsymbol{N_{\rm gal}=}81$
		  & $\boldsymbol{m_{\mathrm{F160W}}^{\rm ref}=}15.30$
		  & $\boldsymbol{\alpha=}0.21$ & $\boldsymbol{\sigma_{\rm LT}^{\rm ref}=}316^{+10}_{-29}$ & $\boldsymbol{\beta=}0.78$ & $\boldsymbol{r_{\rm cut}^{\rm ref}=}6.0^{+4.2}_{-1.6}$ & $\boldsymbol{\gamma=}0.20$  & -- & --\\
		  
		  \hline
		 
	\end{tabular}

	\label{table:inout_lensing}

\end{table*}

\section{Results and discussion} \label{sec:results} 
Section~\ref{sec:BFmodel} presents the results of the six lens models we developed and the rigorous selection of the best-fitting lens model of MACS~J0138.
Subsequently, we focus on the best-fitting \texttt{reference} lens model that we used to study the total mass distribution of the deflector in \Sec~\ref{sec:massdistribution} and the reconstructed surface-brightness distribution of the SNe host galaxy in \Sec~\ref{sec:SBreconstruction}.
We assess in \Sec~\ref{sec:forecasts} the robustness of our new lens model in predicting the positions and magnifications of the observed and future multiple images of SN~Encore and SN~Requiem and we discuss possible implications for future cosmological applications, while \Sec~\ref{ENV} is dedicated to the analysis of the effect of the lens environment.

\subsection{Best-fitting lens model} \label{sec:BFmodel}
The goodness of the lens models presented in \Sec\,\ref{sec:models} in reproducing the observed positions of the multiple images is summarised in \Tab\,\ref{tab:model_stat}, where the values of the statistical estimators introduced in \Sec\,\ref{sec:SLmethod} are given. We recall that lower values of $\chi^2$, rms, BIC, and AIC, and higher values of $\mathcal{L}$ and $\log{E}$ identify the cluster total mass reconstructions that are preferred based on the observables considered in the models.
The model that reproduces the observed positions of the multiple images best is \texttt{ellipticalBCGout}, with a value of the rms of $0\arcsec.22$. When the values of all the figures of merit are considered as well, the three models, including FG, yield similar BIC values, and the \texttt{reference} model is weakly preferred over \texttt{ellipticalBCGout} \citep[][]{Raftery1995}.
Thus, adopting calibrated Faber-Jackson relations with or without the BCG provide consistent results, and the better reproduction of the observed positions of the radial multiples images of systems 5 and 6 is driven by the introduction of an elliptical dPIE mass density profile for the BCG in the \texttt{ellipticalBCGout} model. These systems contribute $\sim 42 \%$ and $\sim 11 \%$ to the total $\chi^2$ term in the \texttt{reference} model, respectively. 
We also remark that it is crucial to include the mass contribution of the foreground spiral perturber FG for an accurate strong-lens model of MACS~J0138. This decreases the minimum $\chi^2$ value by a factor $\sim$3.8.
This comprehensive analysis shows that the \texttt{reference} lens model is statistically preferred while avoiding over-fitting. The lens models we developed after the unblinding between lens modellers in the comparison challenge (Suyu et al. in prep.) show that the inclusion of the massive clump significantly improves the $\chi^2$, by a factor of $\sim2.5$ with respect to the \texttt{reference} model. 
We focus in the following on the results from the \texttt{reference} lens model, which we submitted to the comparison study by Suyu et al. (in prep.). We further discuss the impact of directly modelling the lens environment in Sect.~\ref{ENV}.

The \texttt{reference} best-fit lens model reproduces the observed positions of the multiple images with a root-mean-square offset of $0\arcsec.36$. Figure \ref{fig:macs0138} shows the comparison between the observed (circles) and model-predicted (crosses) positions of the 23 multiple images included in the analysis and the critical lines at the redshift of the strongly lensed SNe ($z_{\rm s1, s2} = 1.949$). 
The observed positions of the two SN multiple image systems and their host galaxy are accurately reproduced, with a remarkable mean precision of only $0\arcsec.05$, which we further discuss in \Sec\,\ref{sec:forecasts}, together with our forecast of the reappearance of future multiple images. We note that some ghost multiple images of systems 1, 2, and 3 are predicted in the centre of the mass density profile associated with the jellyfish galaxy JF-2, which is angularly close to the BCG. Because of the uncertain total mass density profiles of jellyfish galaxies and model degeneracies between the different mass components (which might explain the high value of the velocity dispersion of JF-2, as given in \Tab\,\ref{table:inout_lensing}), we do not consider these multiple images in the following. 
The other multiple image systems are well reproduced by our best-fit \texttt{reference} lens model, which predicts some additional multiple images. In particular, the multiple image 4.2b is predicted to lie close in projection to the jellyfish cluster member JF-2 (orange cross in \Fig\,\ref{fig:macs0138}). The light contamination by this star-forming galaxy both in the JWST and VLT/MUSE data make it extremely challenging to observe this source. Counter images of the systems 4.3 are also predicted close to the locations of the observed multiple image systems 4.1 and 4.2.
Finally, our best-fit model predicts a counter-image of the radial image 6b. While there is some tentative signal in the VLT/MUSE cube, the diffuse emission, together with the averaged seeing of $0\arcsec.8$, prevented us from securely identifying it. 

The previous lens model of MACS~J0138 by \citet{Rodney2021}, also developed with \texttt{Lenstool}, achieved a smaller root-mean-square offset of rms=$0\arcsec.15$ than the offset from all of the models explored in this work (see \Tab\,\ref{tab:model_stat}). This can be explained by the fact that because they exploited shallower VLT/MUSE observations, \citet{Rodney2021} did not include systems 5 and 6, which dominate the contribution to the total $\chi^2$ term. The improvement of our new lens model over those presented by \citet{Newman2018a} and \citet{Rodney2021} is further discussed in the following sections. It further supports the statement that considering the value of the rms alone may not fully quantify the goodness of a given lens model \citep[see also e.g.][]{Acebron2017, Bergamini2023a}.

\subsection{Total mass distribution} \label{sec:massdistribution}
The resulting median values of the parameters of the \texttt{reference} model and the associated $1\sigma$ statistical errors are summarised in \Tab\,\ref{table:inout_lensing} (bottom). 
We note that given our single-plane lens model approximation, the model-predicted velocity dispersion of the background galaxy BG, $\sigma_{0, \rm BG}= 205 \pm 30 ~ \rm km ~s^{-1}$ (see \Tab\,\ref{table:inout_lensing}), agrees to within 1$\sigma$, with the measured stellar kinematics presented by \citetalias{Granata2025} of $\sigma_{0, \rm BG}=237.7 \pm 1.8 ~ \rm km ~s^{-1}$. 

In the left panel of \Fig\,\ref{fig:massdist_SNEncore}, we compare the cumulative projected total mass profile of MACS~J0138 for the six lens models we explored. Regardless of the adopted modelling choices, the cumulative projected total mass profile is robustly determined. The typical statistical uncertainty is  $\sim$ 2.5(6.0)\% at $\sim$ 10(200) kpc and from the BCG. The systematic uncertainties dominate the total error budget, as expected, with a median value of $\sim$ 7.0(18.0)\%. The agreement is particularly remarkable between $\sim$ 20 kpc and $\sim$ 70 kpc from the BCG, where the median projected distance of the 23 multiple images from the BCG is 62 kpc. At these projected distances from the BCG, the statistical plus systematic error is small, $\lesssim$ 3\%.
Based on our best-fitting \texttt{reference} lens model and not taking into account the two line-of-sight galaxies, we measured a precise projected total cluster mass of $M_{\rm tot, SL}(<60 ~\rm{kpc})~=~2.89_{-0.03}^{+0.04}~\times 10^{13} M_{\odot}$.

Owing to the bright emission of the ICM in MACS~J0138, we can compare its total mass distribution measured from our strong-lensing analysis with that inferred from the X-ray emission.
To do this, we performed a spatially resolved spectral analysis of the X-ray data, optimally selecting five circular bins with a maximum radius of 172 arcsec (or $\sim 830$ kpc at $z=0.336$) centred on the BCG position. We find a clear indication of a cool core, despite the large uncertainties on the temperature due to the limited signal and the high temperatures involved. 
We measured $kT=(4.00\pm 0.25) $ keV within 25 kpc and temperatures between 6 and 7 keV beyond 25 kpc. This agrees with the average global temperature found by \citet{Flowers2024}. The presence of a cool core is consistent with the smooth and spherically symmetric surface brightness of the ICM.

After computing the deprojected temperature and electron density profiles, 
we derived the cumulative total mass radial profile of MACS~J0138 
by applying the hydrostatic equilibrium equation. We found a total 
virial mass of $M_{500}= 5.7_{-2.1}^{+3.4}\times 10^{14} M_{\odot}$ 
within $R_{500} =1070_{-150}^{+200}$ kpc. 
Finally, we derived the projected total mass profile as a function of the projected radius.
The latter (in dark red) is compared with those obtained from the strong-lensing analysis in the left panel of \Fig\,\ref{fig:massdist_SNEncore}. The agreement within the $1 \sigma$ uncertainties between the Chandra hydrostatic equilibrium and the strong-lensing cumulative projected mass profiles is very good. This further demonstrates the robustness of our strong-lensing analysis of this unique lens cluster. We estimated a projected hydrostatic mass of $M_{\rm tot, Chandra}(<60 ~\rm{kpc}) = 3.3_{-0.6}^{+0.7} \times 10^{13} M_{\odot}$, which agrees well with the strong-lensing measurement reported above.

The total surface mass density distribution of MACS~J0138 is shown instead (with green contours) in the right panel of \Fig\,\ref{fig:massdist_SNEncore}, where it is compared with the X-ray surface brightness distribution obtained from the Chandra observations. As illustrated, the distribution of the total mass and the mass associated with the ICM as traced from our strong-lensing analysis and the Chandra observations shows that the galaxy cluster is relaxed, with a symmetric mass distribution.
The diffuse dark matter component is very well aligned with the position of the BCG, with an average projected distance of $<10$ kpc in our \texttt{reference} lens model.

\begin{figure}
        \centering
        \includegraphics[scale=0.42]{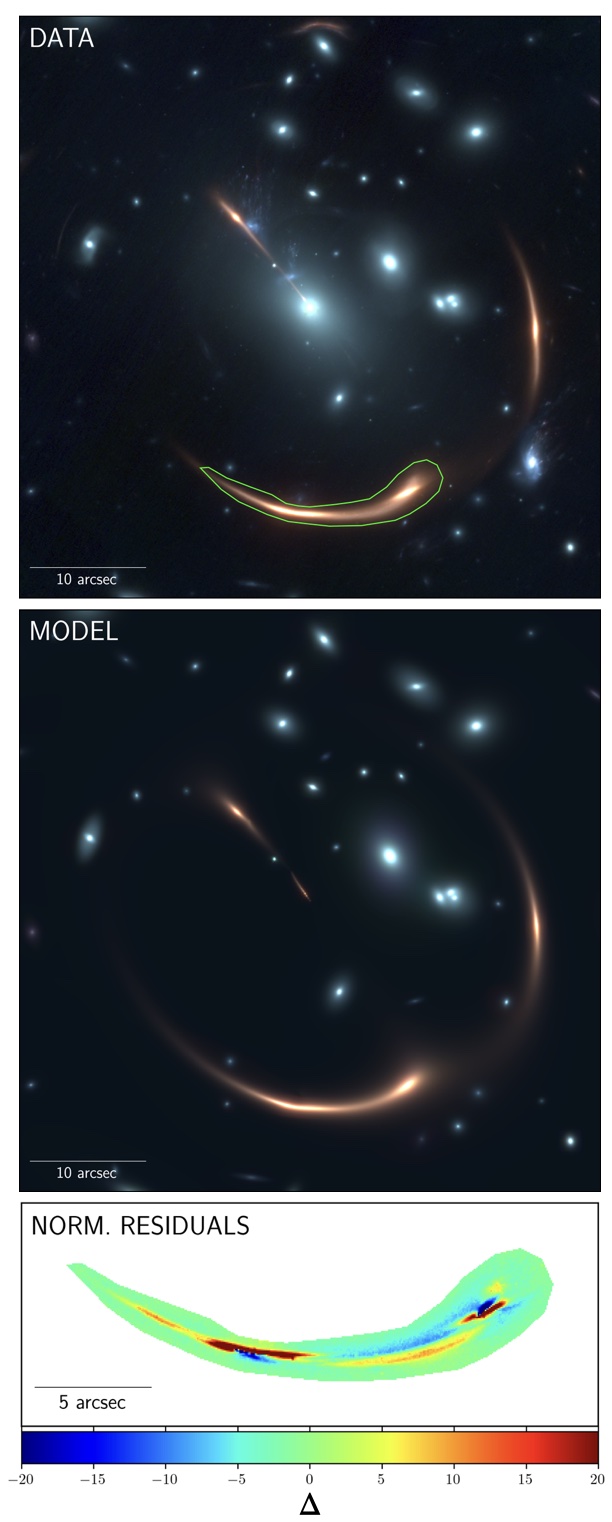}
        \caption{Surface-brightness distribution of MRG-M0138. \textbf{Top and middle panels:} Observed (top panel) and model-predicted (middle panel) surface-brightness distribution of SNe~Requiem and Encore host galaxy, MRG-M0138, from our best-fitting \texttt{reference} lens model. The colour images were created through a combination of three JWST filters (F115W, F200W, and F444W) with a pixel size  of $0.\arcsec 04$. The green region encloses the pixels used in the forward modelling optimisation. \textbf{Bottom panel:} Normalised residuals in the F115W band within the green region shown in the top panel in a range between $-20\sigma$ and $20\sigma$.}
        \label{fig:SB}
\end{figure}

\begin{figure*}
        \centering
        \includegraphics[scale=0.62]{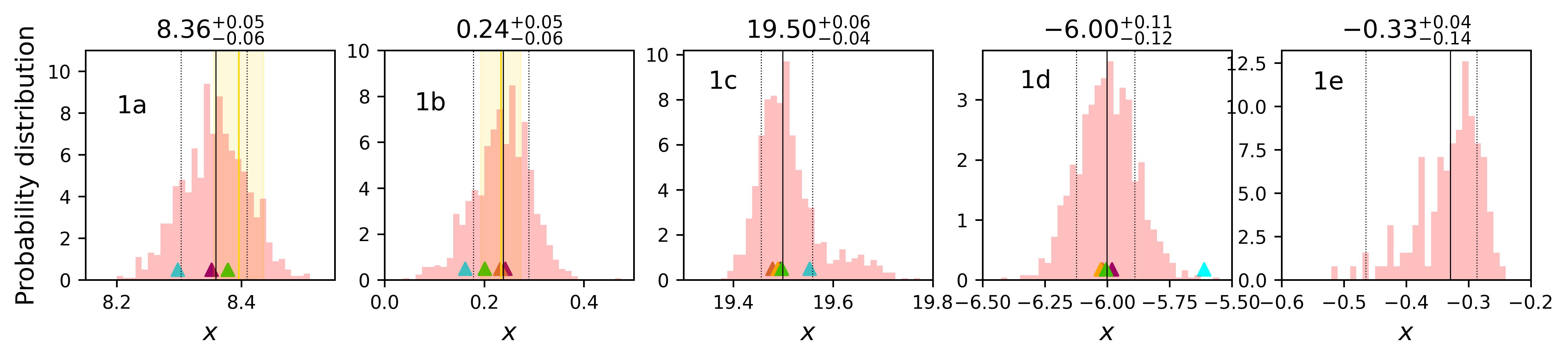}\\
        \includegraphics[scale=0.62]{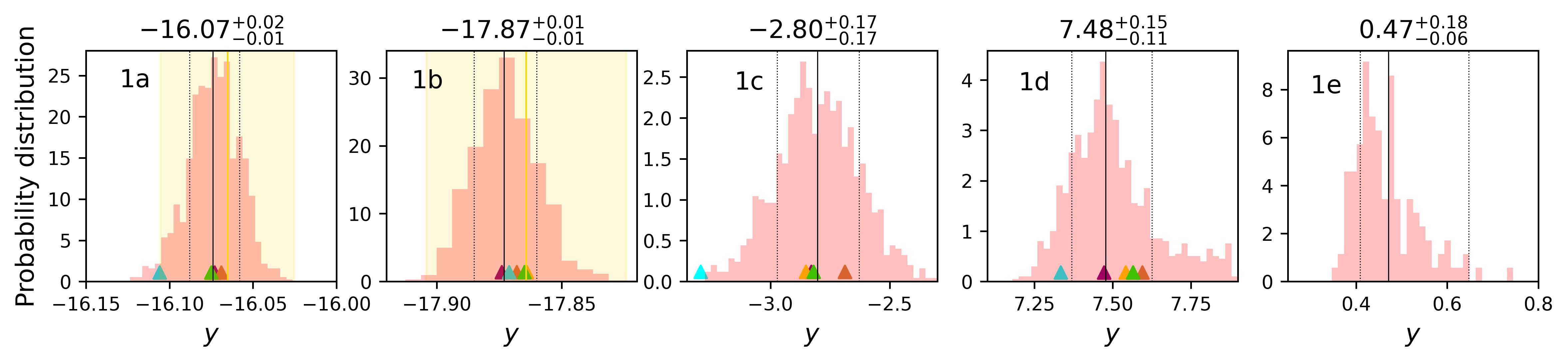}\\
        \includegraphics[scale=0.62]{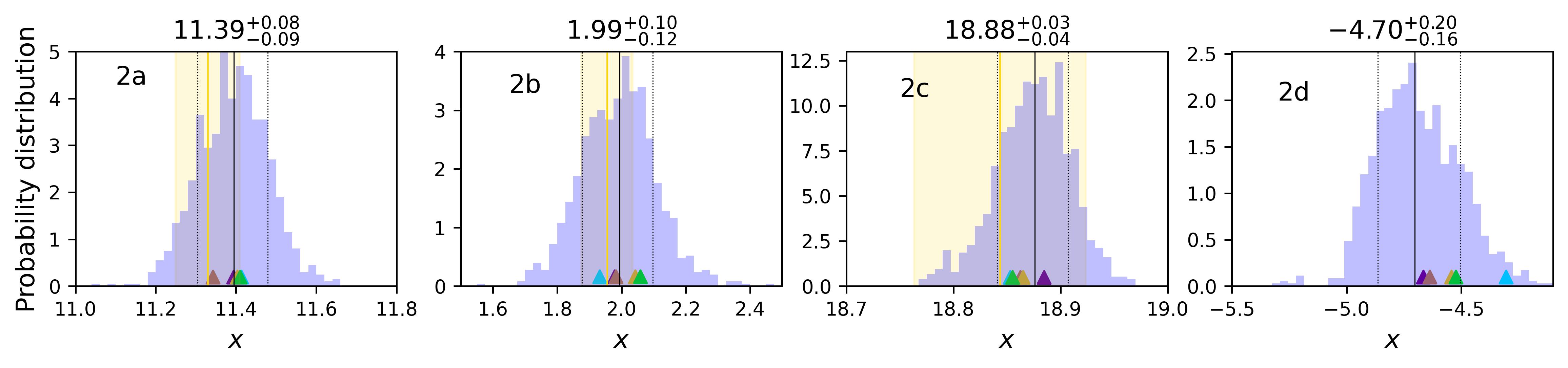}\\
        \includegraphics[scale=0.62]{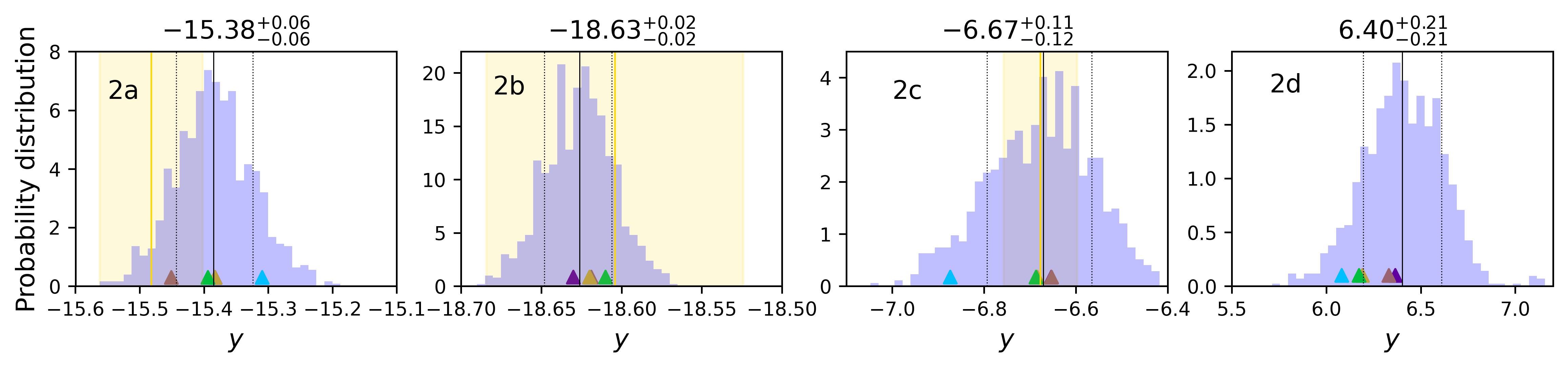}\\
        \caption{Probability distribution functions for the positions of the observed and model-predicted multiples images of SN~Encore (system 1, top) and SN~Requiem (system 2, bottom), in arcsec with respect to our reference position (associated with the BCG). The filled histograms correspond to the marginalised probability distributions obtained from the best-fitting \texttt{reference} lens model, from 1000 different models randomly extracted from the MCMC chain. The solid and dashed black lines correspond to the 50th and the 16th and 84th percentiles of the marginalised distributions, and the corresponding values are reported at the top of each panel. The gold vertical lines in some of the panels denote the observed positions of the multiple images, and the shaded region shows the circularised positional uncertainties. The predicted median values from the lens models \texttt{circularBCGout}, \texttt{ellipticalBCGout}, \texttt{reference-noFG}, \texttt{reference-subs}, and \texttt{subs-noES} are shown with purple, green, cyan, gold, and lime triangles, respectively. 
        }
        \label{fig:histo_xy}
\end{figure*}
\begin{figure*}
        \centering
        \includegraphics[scale=0.62]{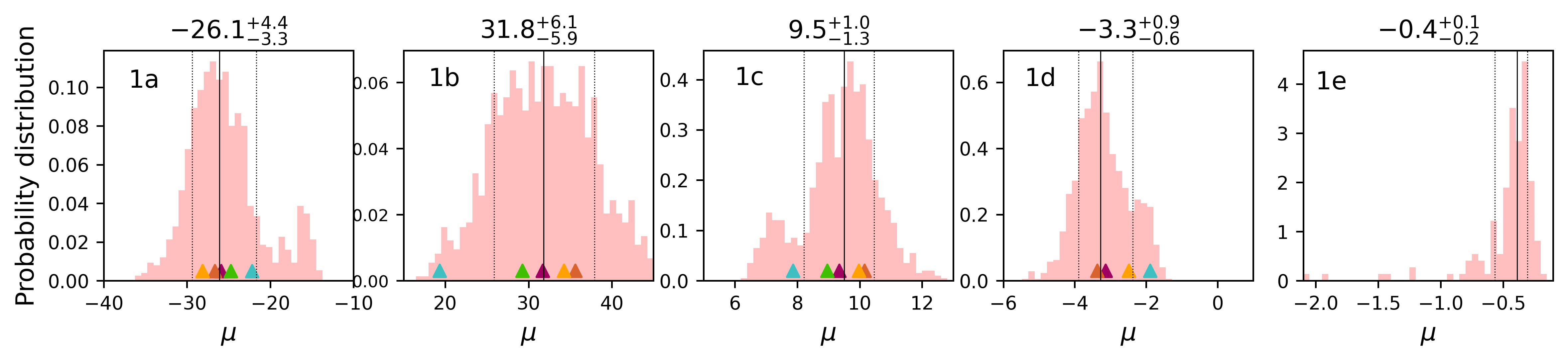}\\
        \includegraphics[scale=0.62]{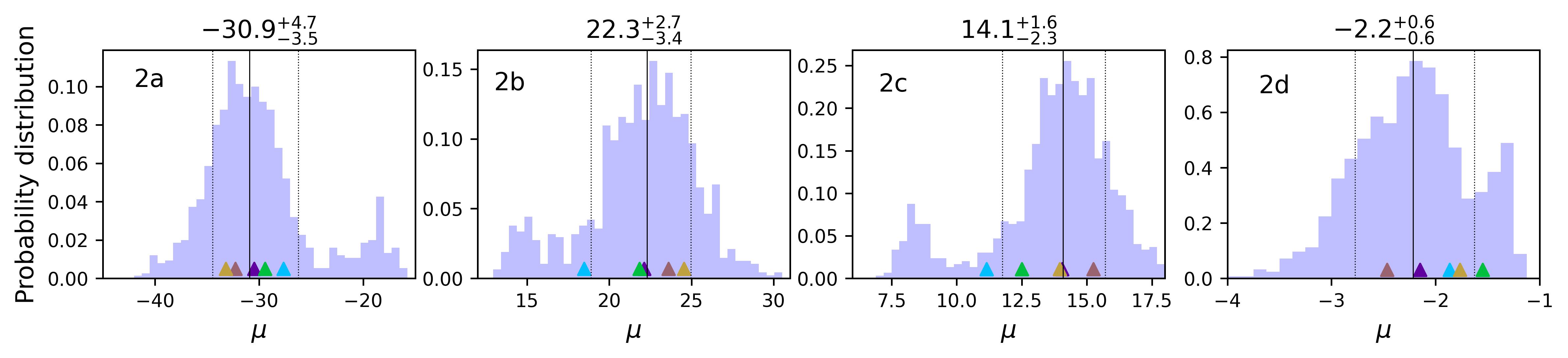}\\
        \caption{Probability distribution functions for the magnification values of the observed and model-predicted multiple images of SN~Encore (system 1, top) and SN~Requiem (system 2, bottom). The shown quantities and labels follow the same notations as in \Fig\,\ref{fig:histo_xy}.
        }
        \label{fig:histo_mu}
\end{figure*} 

\subsection{Reconstruction of the surface-brightness distribution of MRG-M0138} \label{sec:SBreconstruction}

We reconstructed the surface-brightness distribution of the host galaxy MRG-M0138, in which SN~Encore and SN~Requiem exploded, with the forward modelling code \texttt{GravityFM} presented by \citet{Bergamini2023a}. 
\texttt{GravityFM} operates in a Bayesian framework to optimise the model parameter values of a given parametric light profile, representing the strongly lensed sources, by minimising the difference between the measured and model-predicted light intensity of each pixel. 
We explored several source models of MRG-M0138: a single or double S\'ersic profile \citep{Sersic1963} with a fixed S\'ersic index $n=4$, and a double S\'ersic profile with fixed S\'ersic indexes of $n=1$ and  $n=4$. The free model parameters associated with the S\'ersic light profile are the centroid coordinates ($x$ and $y$), the total flux ($F$), the effective radius ($R_{\rm e}$), the axis ratio ($q$), and the position angle ($\theta$).

We selected the pixels associated with the multiple images 3a and 3b for the optimisation. They are encompassed within the green region in \Fig\,\ref{fig:SB} (top panel) to reproduce the more complex morphology of the multiple image 1a (see \Fig\,\ref{fig:macs0138}). These observables (a total of 12113 JWST pixels with a pixel size of $0.\arcsec 04$) were then exploited to predict the best-fit parameter values of the background source.
Based on our best-fitting lens model \texttt{reference}, we predicted a posteriori the surface-brightness distribution of the other counter-images of the host galaxy, that is, without directly optimising it in the lens modelling.
To reconstruct the MRG-M0138 colour surface brightness, we performed the forward modelling in three different JWST filters (F115W, F200W, and F444W), with their associated error maps.

We found that a single S\'ersic profile or a combination of two (with $n=4$, or with $n=4$ and $n=1$) yields a similar fit to the data, and we therefore chose the single S\'ersic as the best-fitting source model in the following because it has fewer (six) free parameters. The goal of this extended reconstruction was to demonstrate the robustness of the lens model and not to estimate the intrinsic properties of the source. We defer these analyses to future extended lens models of MACS~J0138, following similar methods as those implemented by \citet{Acebron2024}.
In \Fig\,\ref{fig:SB} we show the comparison between the observed (top panel) and the model-predicted (middle panel) surface-brightness distribution of the multiple images of MRG-M0138 from our best-fit \texttt{reference} model together with the normalised residuals in the F115W band in a range between $-20\sigma$ and $20\sigma$ (bottom panel). The surface-brightness models of the other galaxies in the field, shown in the middle panel, were taken from \citet{Ertl2025}, as described in \Sec\,\ref{sec:data}. We did not include the light contribution from the BCG to allow for a better visualisation of model-predicted surface-brightness of the radial arc of MRG-M0138, however. 
The surface-brightness of the multiple images of MRG-M0138 is well reproduced. This further demonstrates the robustness of our new lens model. 
The normalised residuals are significant around the multiple image 3a, and our simple source models are not flexible enough to capture the asymmetries and the tilt in the light of MRG-M0138. This limitation might be resolved with surface-brightness models performed on a pixel grid \citep[e.g.][]{Suyu2006}.
We also note that based on the normalised residuals around the foreground galaxy FG, our \texttt{reference} lens model seems to overestimate its total mass, with a best-fit Einstein radius $\theta_{\rm E, \infty}=0.\arcsec13$. The imperfect reconstruction of the radial multiple images (3d and 3e) can be explained by the uncertain total mass density profiles of the jellyfish galaxies, in particular, those of JF-1 and JF-2.
As shown by \citet{Acebron2024}, a direct modelling of the surface-brightness distribution of MRG-M0138 will significantly improve the overall reconstruction and help us to further refine the total mass estimate of angularly close haloes along the line of sight, such as the foreground galaxy FG, or at the lens redshift, such as JF-1, JF-2, and JF-3.

\subsection{Robustness of the model-predicted positions and magnifications of the two strongly lensed supernovae} \label{sec:forecasts}

We show in Figures \ref{fig:histo_xy} and \ref{fig:histo_mu} the probability distribution functions, estimated from 1000 different model realisations randomly extracted from the MCMC chains, of the model-predicted values of the position ($x$ and $y$), and the magnification, $\mu$ of the observed (1a, 1b and 2a, 2b, and 2c), and the future (1d, 1e, and 2d) multiple images of SN~Encore and SN~Requiem, respectively. The magnification values were computed at the model-predicted positions of the multiple images. We highlight the median values obtained and the 68\% confidence level intervals with vertical lines, whilst the gold shaded regions denote the observed positions of the multiple images, considering their circularised positional uncertainties (see \Sec\,\ref{sec:SLmodelling}). The filled histograms correspond to the model-predicted quantities yielded by the \texttt{reference} lens model, and the purple, brown, cyan, gold, and lime triangles represent the model-predicted median values from the \texttt{circularBCGout}, \texttt{ellipticalBCGout}, and \texttt{reference-noFG}, \texttt{reference-subs},  and \texttt{subs-noES} model predictions, respectively.

As anticipated, our lens modelling predicts the reappearance of both SN~Encore and SN~Requiem in the future, with two and one additional multiple image(s), respectively. The multiple images 1d and 2d are expected in the radial arc of MRG-M0138 at RA, Dec=($-6\arcsec.00_{-0\arcsec.12}^{+0\arcsec.11},~7\arcsec.48_{-0\arcsec.11}^{+0\arcsec.15}$) and ($-4\arcsec.70_{-0\arcsec.16}^{+0\arcsec.20},~6\arcsec.40_{-0\arcsec.21}^{+0\arcsec.21}$) from the BCG, respectively (see also the coloured crosses in \Fig\,\ref{fig:macs0138}). Our model predicts (in $15\%$ of the chains) a fifth and central multiple image of SN~Encore, 1e, at RA, Dec=($-0\arcsec.33_{-0\arcsec.14}^{+0\arcsec.04},~0\arcsec.47_{-0\arcsec.06}^{+0\arcsec.18}$). We note that the other three total mass model parametrisations do not predict the appearance of 1e. With a magnification factor of $\mu_{\rm 1e}=-0.4^{-0.1}_{+0.2}$, the observation of this demagnified multiple image, which is angularly close to the bright BCG, would be challenging.

Accurate magnification maps are fundamental for characterising the intrinsic properties of background-lensed sources and to photometrically classify the strongly lensed SN~Requiem \citep{Rodney2021}.
We assessed the impact of systematic uncertainties that are associated with the different total mass parametrisations from \Tab\,\ref{tab:model_stat} by comparing the median values of the model-predicted magnifications, as shown in \Fig\,\ref{fig:histo_mu}. We note that all the models in which the foreground galaxy FG is included provide consistent values, generally, well within the $1\sigma$ interval from the \texttt{reference} model. This entails that given the current observables, the systematic uncertainties arising from different mass modelling choices are comparable to the statistical errors. On the other hand, this analysis also demonstrates that the mass contribution from FG should be considered for accurate and precise magnification estimates, and thus, of the intrinsic properties of MRG-M0138 and SNe~Requiem and Encore.  
Our model-predicted magnification values of the multiple images of SN~Requiem strongly disagree with those presented by \citet{Newman2018a} and particularly with those from \citet{Rodney2021}, however. In detail, the magnification values we estimated are systematically higher than those by \citet{Newman2018a} and \citet{Rodney2021}, by factors ranging from $\sim$1.8 to $\sim$3.0 and from $\sim$3.8 to $\sim$5.5, respectively. The discrepancy might partly be ascribed to the addition of new observables that were revealed in the new JWST and VLT/MUSE observations \citep[systems 1, 5, and 6, see][]{Ertl2025}, and to the improved selection of cluster members \citep[\citetalias{Granata2025},][]{Ertl2025}.
The conflicting model-predicted magnification values are, however, mainly due to the different lens modelling choices, even when the same modelling code is used.
We verified that the model-predicted magnification values by our \texttt{reference} model are robust by removing the multiply lensed systems 1, 5, and 6 from our catalogue of observables, that is, we included the same observables as \citet{Rodney2021}. In particular, we found that the magnification values of the SN~Requiem multiple images are perfectly consistent with those given in \Fig\,\ref{fig:histo_mu}. This further demonstrates that lens modelling choices are important and significantly impact the magnification estimates, especially in the high-magnification regime \citep[][Suyu et al. in prep.]{Treu2016, Meneghetti2017}.
For example, the value of the normalisation of the Faber-Jackson and the truncation radius scaling relations (see \Eq\,\ref{eqSR}) was fixed to given arbitrary values by \citet{Rodney2021}, while in this analysis, the sub-halo mass component was calibrated based on the stellar kinematic measurements of a subset of cluster members (\citetalias{Granata2025}). This allowed us to reduce key model degeneracies between the BCG and the large-scale dark matter mass components (see \Tab\,\ref{table:inout_lensing} and \Fig\,\ref{fig:sigmaSNEncore}), thus building a more accurate total mass model.

\subsection{Impact of the lens cluster environment} \label{ENV}
\begin{figure}[h!]
        \centering
        \includegraphics[width=\columnwidth]{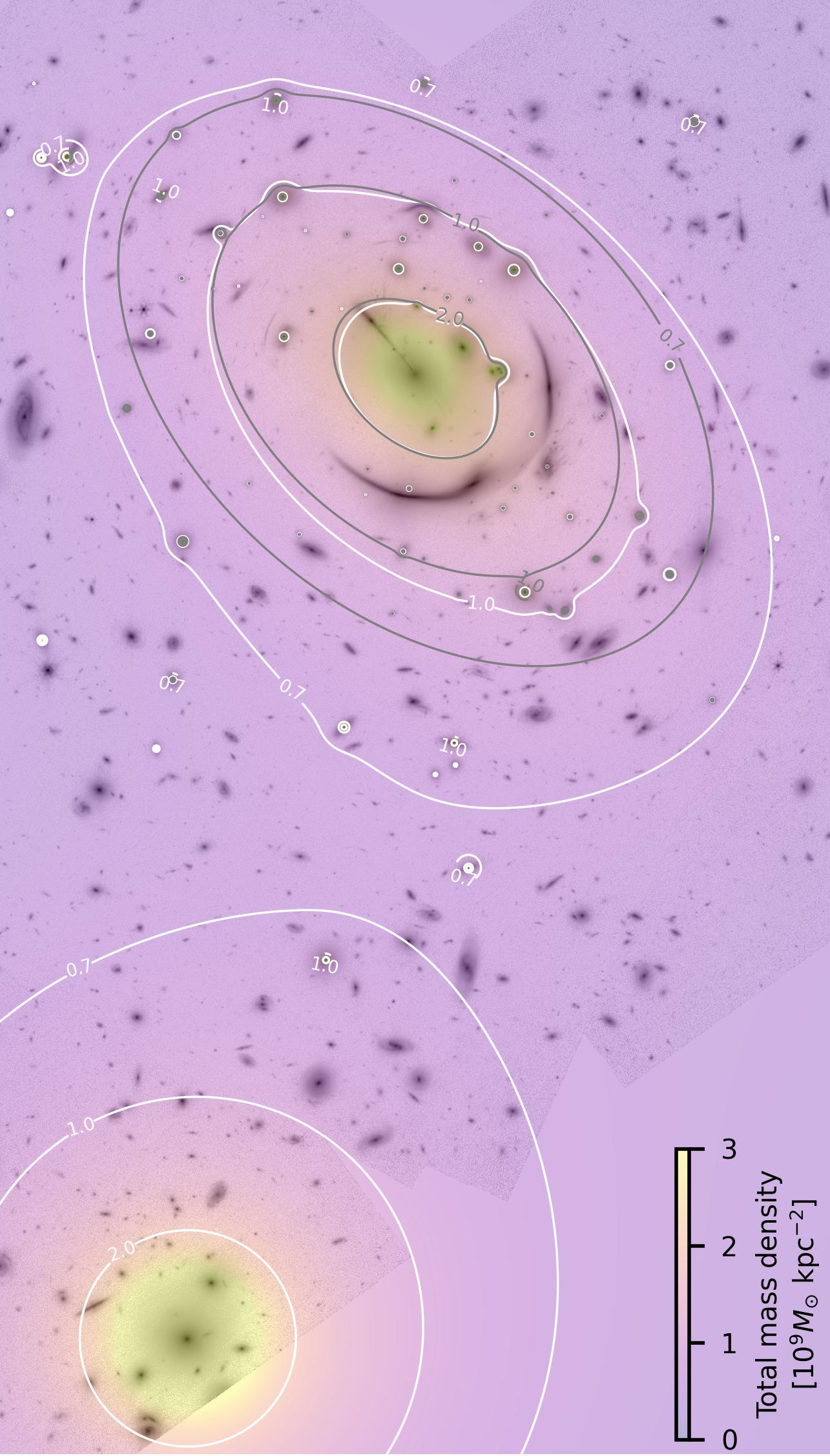}
        \caption{Projected total mass density distribution of the galaxy cluster MACS~J0138 obtained from the best-fit \texttt{reference-subs} model, in units of $10
        ^{12} M_{\odot} ~\rm kpc^{-2}$, overlaid on the JWST/F444W image. The isodensity contours are plotted in grey and white for the \texttt{reference} and \texttt{reference-subs} models, respectively.}
        \label{fig:macs0138_wf}
\end{figure}

In this section, we study the impact of directly modelling the massive substructure $\sim140\arcsec$ south-east of the BCG (see  \Fig~\ref{fig:macs0138_wf}) on the reconstruction of the total mass distribution of the lens, and thus, on the model-predicted positions and magnifications of SN~Requiem and SN~Encore. Based on the current JWST and Chandra data, no clear strong-lensing features are identified and no X-ray emission is detected.

In \Fig~\ref{fig:macs0138_wf} we show the JWST/F444W imaging on which the projected total mass density distribution of the galaxy cluster MACS~J0138, as derived from the best-fit \texttt{reference-subs} model, is overlaid. The figure highlights that the total mass contribution from the distant substructure is comparable to that of the main cluster core, characterised by a high velocity dispersion value, $\sigma_{\rm LT}^{\rm SIS}=725^{+138}_{-262} ~\rm km~s^{-1}$, but with a large statistical uncertainty. The SIS mass density profile represents a simplified parametrisation of the total mass contribution of this distant substructure, as the value of the total mass density diverges in the innermost region of this mass density profile. The comparison of the isodensity contours between the \texttt{reference} and \texttt{reference-subs} models reveals that this external substructure perturbs the elliptical total mass distribution in our \texttt{reference} model, which more closely resembles that found by \citet[][see their \Fig~9]{Ertl2025}.

The \texttt{subs-noES} model yields a poorer fit to the observables than \texttt{reference-subs} (see Table~\ref{tab:model_stat}). This demonstrates that the external shear component in \texttt{reference-subs} (with $\gamma_{\rm ext}=0.04^{+0.03}_{-0.03}$) improves the reproduction of the observed positions of the 23 multiple images in the core of MACS~J0138. We find that its magnitude in the \texttt{reference} lens model ($\gamma_{\rm ext}=0.11^{+0.02}_{-0.02}$; see Table~\ref{table:inout_lensing}) can then only partly be ascribed to the lens environment. Based on the expression of the shear introduced by an SIS mass model at an angular separation $\theta$, that is, $\gamma_{\rm SIS}(\theta)=\theta_{\rm E}/2\theta$, we obtain a value of $\sim0.08$ for our \texttt{reference-subs} model. 
Furthermore, the orientation of the external shear component in \texttt{reference-subs} varies from that in \texttt{reference}, with $\phi_{\rm ext}=122.4\degree$$^{+27.5\degree}_{-33.9\degree}$, indicating a reconfiguration of the external shear field when explicitly modelling the environment.
The residual external shear term required by \texttt{reference-subs} likely reflects complexities in the total mass distribution of the lens that cannot be fully captured with a single elliptical mass component in our parametric modelling or our simple approximation of the galaxy cluster environment (with an SIS mass density profile for both the dark matter and galaxies).

Figures~\ref{fig:histo_xy} and \ref{fig:histo_mu} illustrate the effect of directly modelling the lens cluster environment on the model-predicted positions and magnifications of SN~Requiem and SN~Encore. The median values are consistent within $1\sigma$ with the \texttt{reference} model, especially for the \texttt{reference-subs} mass parametrisation.

While the impact of the lens environment can vary significantly from one field to the next, our analysis demonstrates that in this lensing system, its effect is significant for a robust and precise lens modelling \citep[see also][]{Acebron2017, McCully2017, Mahler2018, Bergamini2023b, Furtak2023}.
Our work proves that future lens models and time-delay cosmography analyses of MACS~J0138, possibly supported with a wide-field spectroscopic follow-up campaign, will highly benefit from directly modelling the mass component associated with the environment of this unique lens system.

\section{Conclusions} \label{sec:conclusions}
We have presented a new and detailed total mass reconstruction of the lens cluster MACS~J0138.0$-$2155, which is a first and crucial step towards future cosmological applications. To do this, we exploited the most recent HST, JWST, and VLT/MUSE observations, prompted by the discovery of SN~Encore. The new observations obtained with the JWST and the VLT/MUSE facilities \citep[see][]{Pierel2024b, Granata2025} have enabled us to construct extended and secure samples of multiple images and cluster members \citep{Ertl2025}. These crucial ingredients in turn yield new and improved total mass models. 
In addition, MACS~J0138.0$-$2155, for which additional multiple images of SN-Requiem and SN~Encore are expected in the future, provides us with a valuable opportunity to blindly test our lens modelling choices and results (Suyu et al. in prep.). 

As one of the seven teams that developed an independent strong-lensing analysis of this unique lens system, we used the parametric software \texttt{Lenstool} \citep{Jullo2007} to model its total mass distribution. We built six lens models by implementing different cluster total mass parametrisations to assess the statistical and systematic uncertainties on the predicted values of the position and magnification of the observed and future multiple images of SN~Requiem and SN~Encore. Our best-fitting \texttt{reference} model includes a single-cored elliptical pseudo-isothermal mass density profile, a sub-halo mass component with 81 cluster members anchored by the VLT/MUSE stellar kinematic information \citep[see also][]{Granata2025}, three jellyfish cluster member galaxies, and one foreground and one background galaxy whose total mass parameters were free to vary. 
The observed positions of the multiple images were reproduced with a root-mean-square offset of $0\arcsec.36$.
We measured a precise projected total cluster mass of $M_{\rm tot, SL}(<60 ~\rm{kpc}) = 2.89 \times 10^{13} M_{\odot}$, with a statistical plus systematic uncertainty of $\lesssim 3\%$ in the region in which multiple images were identified. In addition, our measurements are consistent with those derived independently from the X-ray analysis of archival data from the Chandra observatory.
We provided the model-predicted values of the position and magnification of the observed and future multiple images of SN~Requiem and SN~Encore. The systematic uncertainties arising from different mass modelling choices are comparable to the statistical errors when the mass contribution from the foreground spiral galaxy FG is considered. Moreover, the inclusion of the massive structure $140\arcsec$ south-east of the BCG has a significant impact on the positions of the observed multiple images in the core of the galaxy cluster. 

Our position-based strong-lens modelling of MACS~J0138.0$-$2155 can further be enhanced in several ways. For instance, by refining the modelling of the lens system environment, by implementing a multi-plane lens analysis \citep[e.g.][]{Chirivi2018}, by directly modelling the surface-brightness distribution of the
supernovae host galaxy MRG-M0138 \citep[e.g.][]{Acebron2024}, by including the hot-gas mass component \citep[e.g.][]{Bonamigo2017, Bonamigo2018}, and by considering the measured time delays (and their uncertainties) between the multiple images of SN~Requiem \citep{Rodney2021} and SN~Encore (Pierel et al. in prep.) as observables. As previously highlighted in other studies \citep[e.g.][]{Grillo2018, Acebron2022}, this method will be fundamental for a precise and accurate measurement of the Hubble constant value with MACS~J0138.0$-$2155. 

\begin{acknowledgements}
We kindly thank the referee for the useful comments received. This research is based partly on observations made with the NASA/ESA Hubble Space Telescope obtained from the Space Telescope Science Institute (associated with the programmes 14496, 15663, and 16264), which is operated by the Association of Universities for Research in Astronomy, Inc., under NASA contract NAS 5–26555, and on observations made with the NASA/ESA/CSA James Webb Space Telescope (associated with the programmes 2345 and 6549). The data were obtained from the Mikulski Archive for Space Telescopes at the Space Telescope Science Institute, which is operated by the Association of Universities for Research in Astronomy, Inc., under NASA contract NAS 5-03127 for JWST. The specific observations used in this work can be accessed via DOI:\href{https://doi.org/10.17909/snj9-an10}{10.17909/snj9-an10}.
AA warmly thanks the P.I.s of the HST, JWST, and VLT/MUSE observations obtained upon the discovery of SN Encore, Andrew Newman, Justin Pierel, and Sherry H. Suyu for making these data available and Marceau Limousin for the useful discussions. 
AA acknowledges financial support through the Beatriz Galindo programme and the project PID2022-138896NB-C51 (MCIU/AEI/MINECO/FEDER, UE), Ministerio de Ciencia, Investigación y Universidades.
PB, PR, MM, GG, and CG acknowledge financial support through grants PRIN-MIUR 2017WSCC32 and 2020SKSTHZ. SE thanks the Max Planck Society for support through the Max Planck Fellowship for SHS. SS has received funding from the European Union’s Horizon 2022 research and innovation programme under the Marie Skłodowska-Curie grant agreement No 101105167 — FASTIDIoUS.
\end{acknowledgements}

%
%

\bibliographystyle{aa}
\bibliography{bibliography}

\begin{appendix}

\section{Summary of the best-fitting lens models of MACS~J0138} \label{A1}
\begin{figure*}
        \centering
        \includegraphics[scale=0.3]{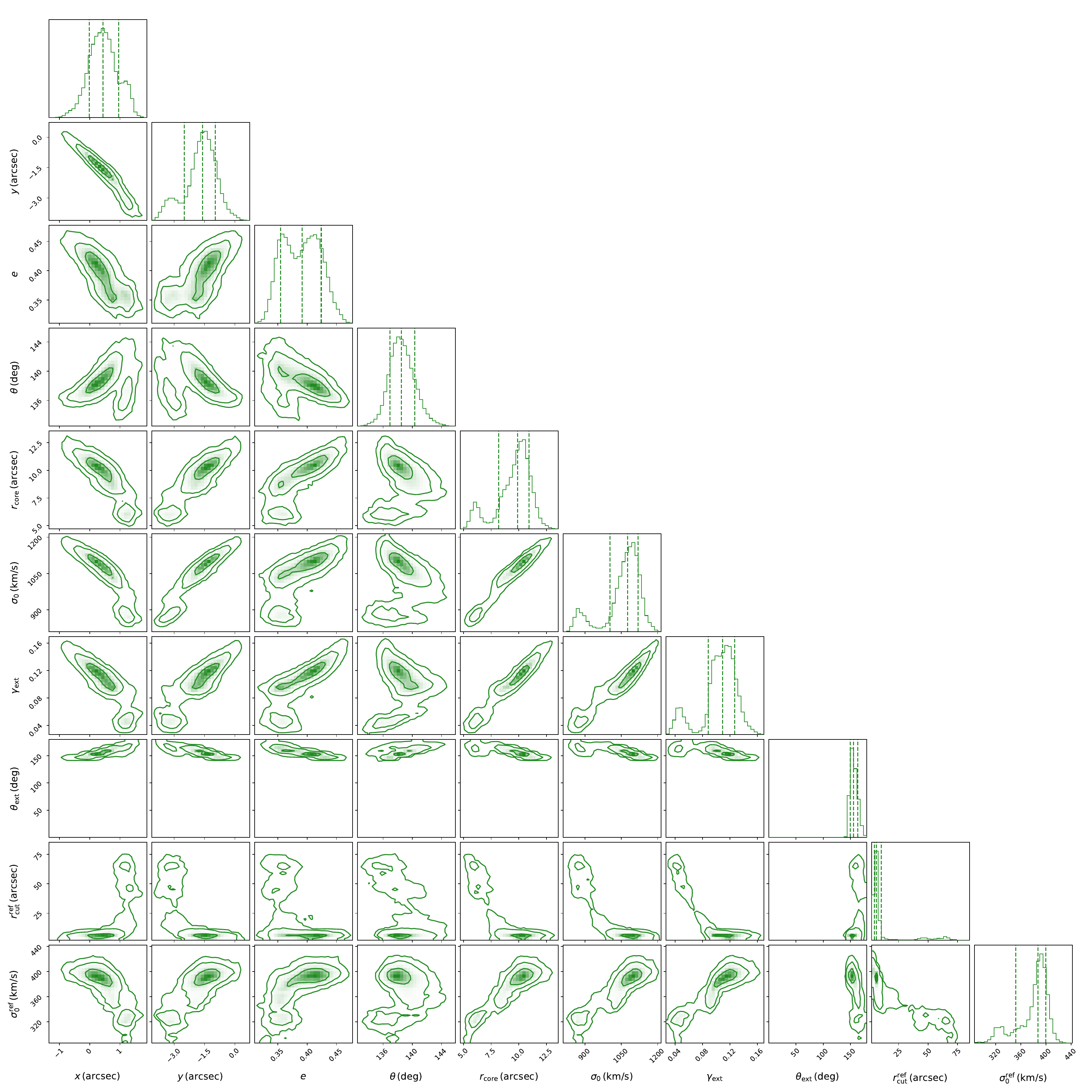}
        \caption{Posterior probability distributions of the parameter values of the cluster-scale and sub-halo mass components for the \texttt{reference} lens model. The contours correspond to the 1 $\sigma$ confidence levels, and the vertical dashed lines in the histograms correspond to the 16th, 50th, and 84th percentiles. 
        }
        \label{fig:posterior}
\end{figure*}

In Appendix \ref{A1}, \Tab~\ref{tab:model_stat} provides the values of several statistical estimators (see \Sec \ref{sec:SLmodelling}) to quantify the goodness of the several cluster total mass models explored in this work. 
We also present, in \Fig\,\ref{fig:posterior}, the posterior probability distributions of the parameter values of the model free parameters for the \texttt{reference} lens model.

\begin{table*}[]
\renewcommand\arraystretch{1.2}
  \setlength\tabcolsep{0.4em}
	\centering          
\caption{Description of the characteristics of the four developed lens models of MACS~J0138 with the parametric pipeline \texttt{Lenstool}. We list the number of model free mass parameters ($N_{\rm  mass~ param}$), where $N_{\rm param}$ = $N_{\rm  mass~ param} + 2\times N_{\rm fam}$, the resulting degrees of freedom ($\nu$), and the values of the different statistical estimators used, as described in \Sec \ref{sec:SLmodelling}. We note that the reported $\chi^{2}$ value corresponds to the the minimum $\chi^{2}$ before rescaling the positional uncertainties of the multiple images. The first four lens models were created during the blind comparison challenge described by Suyu et al. (in prep.) while the two last ones were developed after the unblinding with the other lens modelling teams.}
\begin{tabular}{|c|c|c|c|c|c|c|c|c|}
\hline
\textbf{Model} & $\mathbf{N_{\rm mass~ param}}$ & $\mathbf{\nu}$  & $\mathbf{\chi^2}$ & \textbf{rms} [$\arcsec$] & $\mathbf{\log{\mathcal{L}}}$ & $\mathbf{\log{E}}$ & \textbf{BIC} & \textbf{AIC} \\
\hline
\texttt{reference} & 20 & 10 & 24.3 & 0.36 & 26.3 & $-$70.4 & 85.3 & 19.4\\
\hline
\texttt{circularBCGout} & 22 & 8 & 19.5 & 0.31 & 28.68 & $-$64.8 & 88.1 & 18.6\\
\hline
 \texttt{ellipticalBCGout} & 24 & 6 & 10.7 & 0.22 & 33.1 & $-$56.2 & 87.0 & 13.8\\
\hline
\texttt{reference-noFG} & 18 & 12 & 90.0 & 0.39 & $-$6.6 & $-$96.4 & 143.4 & 81.2 \\
\hline
\hline
\texttt{reference-subs} & 21 & 9 & 9.8 & 0.35 & 14.9 & $-80.6$ & 111.9 & 44.2  \\
\hline
\texttt{subs-noES} & 19 & 11 & 15.1 & 0.41 & 12.3 & $-46.3$ & 109.4 &  45.4\\
\hline
\end{tabular}
\label{tab:model_stat}
\end{table*}

\end{appendix}

\end{document}